\documentstyle[12pt,aasms4]{article}

\newcommand{\Msun}{\mbox{ M}_{\odot}}
\newcommand{\gae}{\mathrel{>\kern-1.0em\lower0.9ex\hbox{$\sim$}}}
\newcommand{\kms}{km~s$^{-1}$}
\newcommand{\NV}{\ion{N}{5}\,$\lambda\lambda\,1238.8,\,1242.8$\ }
\newcommand{\CIV}{\ion{C}{4}\,$\lambda\lambda\,1548,\,1550$\ }

\lefthead{Boroson et al.}
\righthead{Models of LMC X-4}

\begin{document}

\title{Models of X-ray Photoionization in LMC X-4: Slices of a Stellar
Wind}

\author{Bram Boroson and Timothy Kallman}
\affil{Goddard Space Flight Center, Greenbelt, MD 20771;
bboroson@falafel.gsfc.nasa.gov, tim@xstar.gsfc.nasa.gov}

\and

\author{Richard McCray}
\affil{JILA, Campus Box 440, University of Colorado, Boulder, CO
80309;
dick@jila.colorado.edu}

\and

\author{S.D. Vrtilek and John Raymond}
\affil{Center for Astrophysics, 60 Garden Street, Cambridge, MA 02138;
svrtilek@cfa.harvard.edu, jraymond@cfa.harvard.edu}


\begin{abstract}

We show that the orbital variation in the UV P~Cygni lines of the X-ray
binary LMC~X-4 results when X-rays photoionize nearly the entire region
outside of the X-ray shadow of the normal star.  We fit models
to Goddard High Resolution Spectrograph (GHRS) observations of \ion{N}{5}
and \ion{C}{4} P~Cygni line profiles.  Analytic methods assuming a
spherically symmetric wind show that the wind
velocity law is well-fit by $v\propto(1-1/r)^\beta$, where $\beta$ is
likely $\approx1.4-1.6$ and definitely $<2.5$.  Escape probability models
can fit the observed P~Cygni profiles, and provide measurements of the
stellar wind parameters.  The fits determine
$L_{x}/\dot{M}=2.6\pm0.1\times10^{43}$~erg~s$^{-1}\Msun^{-1}$~yr,
where $L_x$ is the X-ray luminosity and $\dot{M}$ is the mass-loss rate of
the star.  Allowing an inhomogeneous wind improves the fits.
IUE spectra show greater P~Cygni absorption during the second half of the
orbit than during the first.  We discuss possible causes of this effect.
\end{abstract}

\section{Introduction}

Theorists have made considerable progress in explaining the stationary
properties of OB star winds, and there is now a consensus that the winds are
accelerated through resonance line scattering of the stellar continuum
(Lucy \&\ Solomon 1970; Castor, Abbott, \&\ Klein 1975).  The spectral 
lines predicted by detailed models match observations for a wide variety of
ionization species (Pauldrach 1987), and for stars across a range of 
metallicity and evolutionary status (Kudritzki, Pauldrach, \&\ Puls,
1987; Pauldrach et al. 1988). 

However, it has become clear that real stellar winds are time-variable and
contain inhomogeneities which, although expected on theoretical grounds,
are difficult for the theory to treat properly.  
There is independent evidence for these inhomogeneities from the black 
absorption troughs (Lucy 1982a) and variable ``discrete absorption features''
seen in wind-formed
P~Cygni lines (Massa et al. 1995), in the X-ray emission from OB 
stars (Harnden et al. 1979;
Seward et al. 1979; Corcoran et al. 1994), and from polarimetric 
observations
(Brown et al. 1995), IR observations (Abbott, Telesco, \&\ Wolff 1984) and
radio observations (Bieging, Abbott, \&\ Churchwell 1989).  Theoretically,
the scattering mechanism thought to be responsible for the acceleration of
the wind should be unstable, leading to shocks (MacGregor, Hartmann, \&\
Raymond 1979; Abbott 1980; Lucy \&\ White 1980; Owocki \& Rybicki 1984). 
Although these instabilities have been modelled numerically,
the models encounter nonlinearities (Owocki, Castor, \&\ Rybicki 1988).

The unknown nature of the wind inhomogeneities has hindered attempts to 
infer from observations such
properties of stellar winds as the mass-loss rates,
which are critical for understanding massive star evolution.  
The observations are, in essence, averages over the entire
nonuniform and time-variable wind, and the unknown weighting of these
averages prevents us from knowing the magnitude of the total
mass outflow.
Overcoming these difficulties is a major goal of current studies of hot
stellar atmospheres.

When the OB star has an accreting compact companion,
X-rays from the compact object can be used as a probe
to derive parameters of the undisturbed wind, such as its mass-loss
rate, expansion velocity, and radial velocity law.
In the ``Hatchett-McCray effect'' (Hatchett \&\ McCray 1977; McCray et 
al. 1984), the X-rays remove the ion
responsible for a P~Cygni line from a portion of the wind, resulting in 
orbital variations of P~Cygni profiles.  This effect has been 
observed in several High Mass X-ray Binaries
(HMXB; Dupree et al. 1980;  Hammerschlag-Hensberge 
et al.
1980; van der Klis, et al. 1982; Kallman \&\ White 1982; Haberl, White \&\
Kallman 1989).  
Analysis of the velocities over
which the P~Cygni lines vary has suggested that in some cases, the stellar
wind radial velocity dependence can be nonmonotonic (Kaper,
Hammerschlag-Hensberge, \&\ van Loon 1993). 
Whether this is intrinsic to the OB star wind or is a result of the
interaction with the compact object is not known.  

LMC~X-4 shows a more pronounced Hatchett-McCray effect than any other
X-ray binary.  The UV P~Cygni lines of \ion{N}{5} at 1240\AA\ appear
strong in low-resolution IUE spectra at $\phi=0$, but nearly disappear at
$\phi=0.5$ (van der Klis et al. 1982). The equivalent width of the
\ion{N}{5} absorption varies by more than 100\% throughout the orbit,
suggesting that the X-ray source may remove \ion{N}{5} from the entire
region of the wind outside of the X-ray shadow of the normal star.

If the X-ray ionization of the wind is indeed this thorough, then the 
change in the P~Cygni profile between two closely 
spaced orbital phases is largely the result of the X-ray shadow of the 
primary advancing through a thin ``slice'' of the stellar wind.  Thus
models of the changing P~Cygni profiles should be sensitive to a region
of the wind that is small in spatial extent.
Vrtilek et al. (1997, Paper~I) obtained 
high-resolution spectra of the \ion{N}{5} lines
(and the \CIV lines, which show weaker
variability) with the Goddard High Resolution Spectrograph (GHRS) aboard 
the Hubble Space Telescope.  These spectra, unlike the earlier 
IUE spectra, provide a high enough resolution and signal to noise ratio 
to allow us to examine the detailed velocity structure of 
the P~Cygni lines over closely spaced intervals of the binary orbit.

LMC X-4 consists of a main-sequence O star (variously classified as O7--9
III--V) of about $15\Msun$ and a $\approx1.4\Msun$ neutron star with a
spin period of 13.5 seconds (Kelley et al. 1983).  Pulse delays from the
neutron star have established the projected semimajor axis of the orbit
precisely as
26~light seconds.  The 1.4~day orbital period is also seen in X-ray
eclipses (Li, Rappaport, \& Epstein 1978) and modulations of the optical
brightness (Heemskerk \&\ van Paradijs 1989).  The X-ray spectrum in the
normal state is hard (photon power-index $\alpha\approx0.8$) with little
absorption of the soft X-rays (Woo et al. 1996 give $N_{\rm
H}=1.14\pm0.005 \times 10^{21}$ cm$^{-2}$).  The total quiescent X-ray
luminosity is $\sim10^{38}$~erg~s$^{-1}$, near the Eddington limit for a
$1.4\Msun$ neutron star. 

X-ray flares, in which the pulse-averaged X-ray luminosity can exceed
10$^{39}$~erg~s$^{-1}$, occur with a frequency of about once a day (Kelley
et al. 1983; Pietsch et al. 1985; Levine et al. 1991). These flares
have been interpreted as dense blobs of wind material crashing down on the
surface of the neutron star (White, Kallman, \&\ Swank 1983; Haberl,
White, \&\ Kallman 1989).  It is an open question whether these blobs
represent the natural state of the stellar wind, or are produced by
interactions between the X-ray source and an unperturbed stellar wind.  An
accreting compact object embedded in a stellar wind can affect the wind
through the heating, ionization, and radiation pressure of its X-ray
emission, and through its gravity.  Numerical simulations (Blondin et al. 
1990, 1991) of the modifications of the stellar wind by the compact object
show that the wind can form accretion wakes and disk-like structures (even
in systems which are not thought to have substantial mass-transfer through
Roche lobe overflow).  Thus an investigation of instabilities in the wind
of LMC~X-4 also bears on the cause of the X-ray flares.

We present a series of increasingly more sophisticated
models of the P~Cygni lines in LMC~X-4, as observed with the GHRS and
reported in Paper~I.  
In \S\ref{sec:compare}, we compare the \ion{N}{5} and
\ion{C}{4} P~Cygni lines to those
of isolated OB stars and to those of other massive X-ray binaries.  
We make inferences from analytic and approximate models to the P~Cygni
line variations in \S\ref{sec:simple}.  Then we attempt more detailed
models in \S\ref{sec:numeric}, and derive
best-fit parameters for the wind and X-ray luminosity.  

We note that this is the first detailed fit of P~Cygni lines that has
been attempted for a HMXB.  The
sophistication of the models that we present is justified by the high
quality of the data on this system that should be obtained in the near
future. 

\section{Comparison spectra\label{sec:compare}}

In this section, we compare the UV line profiles of LMC~X-4 with the line
profiles of similar stars.  

\subsection{Other X-ray binaries}

The strength of the orbital variation in P~Cygni lines (the
Hatchett-McCray effect) differs among the HMXB.  Figure~1 compares the
\ion{N}{5} lines at two orbital phases with the corresponding features
in
4U1700-37 and Vela~X-1 (the latter two objects were observed with IUE at
high spectral resolution in the echelle mode).  The optical companion of
4U1700-37 is very massive ($\sim50\Msun$) and hot (spectral type O6.5
with effective temperature 42,000K; Kaper et al. 1993).  In this system
the \ion{N}{5} resonance line is saturated and shows little or no
variation over
orbital phase; this saturation is caused by the very strong stellar wind
produced by the extremely massive and hot companion.  The companion of
Vela~X-1 has a lower mass (23$\Msun$) and is considerably cooler (spectral
type B0.5 with an effective temperature of 25,000K; Kaper et al. 1993).
The wind in LMC~X-4 is expected to be weaker than in Vela X-1 and
4U1700-37, as the optical companion is not a supergiant.  As the
abundances of metals are lower in the LMC than in the Galaxy,
lower mass-loss rates for a given spectral type are expected.  The low
abundances will also make the \ion{N}{5} lines less saturated and the
effects of
X-ray ionization more visible.  The
X-ray luminosity of LMC~X-4 is the highest of the three systems.  As a
result of these considerations, it is not surprising that the effects of
X-ray ionization on the \ion{N}{5} lines are strongest in
LMC~X-4.  For all three systems the \ion{C}{4} resonance lines 
show little orbital variation during the phases observed.

\subsection{Photospheric lines at $\phi\sim0.5$ \label{sec:photo}}

The effects of the X-ray photoionization on the blueshifted P~Cygni
absorption are greatest for LMC~X-4 near $\phi=0.5$.  The observed
\ion{N}{5}
line profiles still show residual absorption at $\phi=0.42$. Paper~I
interpreted this absorption to be the photospheric \ion{N}{5} doublet,
which is
masked by the wind absorption during X-ray eclipse.  If this
interpretation is correct, the width of the absorption is determined by
the rotation of the normal star, not the wind expansion velocity. 
Alternately, at the base of the wind where the expansion velocity is low,
the wind may be dense enough that \ion{N}{5} can still exist even in the
presence of X-ray illumination.

Nearly all early-type stars whose \ion{N}{5} photospheric absorption lines
could
be usefully compared with those of LMC~X-4 also have strong wind
absorption.  The O9~IV star $\mu$Col (HD~38666) seems to be a promising
candidate for comparison, as its \ion{N}{5} line has negligible
red-shifted
emission, and the blue-shifted absorption is confined to features near the
rest wavelength.  

Although this is the best candidate for the intrinsic \ion{N}{5}
photospheric absorption spectrum that we were able to obtain, Wollaert,
Lamers, \&\ de Jager (1988) suggest that the \ion{N}{5} and \ion{C}{4}
lines in this star are wind-formed because they are asymmetric.  We have
compared IUE and HST spectra of $\mu$Col with our spectra of LMC~X-4, and
although the detailed fit is poor (Figure~2a), the strength of the
absorption in $\mu$Col is comparable to the absorption in LMC~X-4 at
$\phi=0.4$.  This provides qualified support that the photospheric
\ion{N}{5} lines of subgiant O stars can be comparable to the lines we see
in LMC~X-4. 

We also compare the LMC~X-4 spectra with the optical absorption lines
which Hutchings et al. (1978) concluded were photospheric.  Hutchings et
al. found that the optical lines were consistent with an orbital velocity
of $50-60$~km~s$^{-1}$ and a systemic velocity of $\approx
280$~km~s$^{-1}$.  They measured the width of the optical
lines to be
$\sim170$~km~s$^{-1}$, which would be expected if the primary rotates
with a period twice as long as the binary orbit. We find a good fit to
the \ion{N}{5} absorption profile using Gaussians with width
(approximately equal to $v \sin i$ width) 170~km~s$^{-1}$ and optical
depths $\tau_{\rm b}=0.6$ and $\tau_{\rm r}=0.3$ in the blue and red
components (Figure~2b).

While the \ion{N}{5} absorption velocity agrees with both the optical
absorption line velocity found by Hutchings et al. and with the systemic
velocity of the LMC (Bomans et al. 1996), the cross-over between
absorption and emission occurs at relative red-shifts of
$100-200$~km~s$^{-1}$.  This may suggest that LMC~X-4 has a red-shift
relative to the LMC and that the ``photospheric'' absorption is actually
blue-shifted P~Cygni absorption.  

In the absence of a clear means of separating the wind and photospheric
absorption, we will allow, in the more detailed analysis that follows, the
strength of a photospheric absorption line to be an adjustable free
parameter.  
We note that 
Groenewegen \&\ Lamers (1989)
fit P~Cygni lines observed with IUE, and that for the stars
HD~24912, 36861, 47839, and 101413 (with spectral types O7.5\,III,O8\,III,
O7\,V, and O8\,V respectively), the best-fit values of $\tau_{\rm r}$
for \ion{N}{5} are 0.0, 0.4, 0.3, and 0.3 respectively.

\section{Simple Models\label{sec:simple}}

In the discussion that follows, we assume that the orbital variation of
the \ion{N}{5} P~Cygni lines is caused
principally by the Hatchett-McCray effect, that is, through the
photoionization of the stellar wind by the X-ray source.  There do not
appear to be any variable Raman-scattered emission lines which Kaper et
al. (1993) suggested cause an orbital variability in the X-ray binary
4U~1700-37.

\subsection{Equivalent width variations (IUE and HST
observations)\label{sec:IUE}}

Following van der Klis et al. (1982), we examine the orbital variation of
the equivalent width of \ion{N}{5} in LMC~X-4 using a simple numeric
model.
Hatchett \&\ McCray (1977) showed that the X-ray ionized zones could be
approximated as surfaces of constant $q$, where each point in the wind has
a value of $q$ given by \begin{equation} \frac{L_{\rm x}}{n_{\rm x} D^2} q
= \xi. \end{equation} Here, $L_{\rm x}$ is the X-ray luminosity, $n_{\rm
x}$ is the electron density at the radius of the X-ray source ($D=1.77$
stellar radii in the case of LMC~X-4), and $\xi$ is the ionization
parameter $\xi\equiv L_{\rm x}/n_{\rm e} r_{\rm x}^2$, where $n_{\rm e}$
is the electron density at the given point, and $r_{\rm x}$ is the
distance from that point to the X-ray source.  
Given a constant mass-loss
rate $\dot{M}$ for the wind, $n_{\rm e}$ can be found from mass
conservation:  \begin{equation} n_{\rm e} = 1.2 \frac{\dot{M}}{4 \pi r^2
v} m_{\rm H}^{-1}.\label{eq:conserve} \end{equation}
We illustrate these parameters in Figure~3.

For large values of $q$, the surfaces of constant ionization are spheres
surrounding the X-ray
source, while for $q<1$, the surfaces are open.  Where the surfaces
extend into the primary's X-ray shadow, a conical shape must be used
instead of the constant $q$ surface.
We model the absorption line variation using the velocity law
$v\propto (1-1/x)$, where $x$ is the wind radial distance in units of the
stellar radius.  According to van der Klis et al. the results are not
sensitive to the velocity law assumed.  We account for the orbital
inclination $i$ ($i=66.2^{+2.5}_{-2.8}\arcdeg$, Pietsch et al. 1985) by
using
a corrected orbital phase $\phi^\prime$ such that $\cos 2\pi
\phi^\prime=\cos 2 \pi \phi \sin i$.  To model the orbital variation of
equivalent width, we used the best-fit escape probability model we
describe in \S\ref{sec:basic}, except that the only X-ray ionization
occurs within the zone bounded by $q$, which is completely ionized.

We measured the equivalent width of \ion{N}{5}, \ion{C}{4}, and
\ion{Si}{4} for all the IUE
observations of LMC~X-4 (using the Final Archive version of the data,
which has been processed by NEWSIPS, the NEW Spectral Image Processing
System, de la Pena et al. 1994). We used the ephemeris of Woo et al.
(1996) to find the corresponding orbital phases for each spectrum.  In
Figure~4, we compare theoretical variations in equivalent width with those
observed with IUE and the GHRS.  That is, we plot
\begin{equation}
\Delta \equiv(W_\lambda-\bar{W_\lambda})/\bar{W_\lambda}
\end{equation}
where $W_\lambda$ is the equivalent width, and $\bar{W_\lambda}$ is the
average value of $W_\lambda$.  Here, $W_\lambda$ measures only the 
equivalent width of the absorption, and not the P~Cygni emission.
As the
GHRS spectra were obtained during less than a complete binary orbit, we
use the average equivalent width from the IUE observations to find the per
cent variation in equivalent width for the GHRS spectra as well. 

The results shown in Figure~4 are consistent with $q\lesssim1$, in which
case the X-ray source photoionizes at least the entire hemisphere opposite
the primary, and possibly as much as the entire wind outside of the
primary's shadow.  The fit to the IUE data is better for $q=1$
($\chi^2=5.5$) than for full X-ray photoionization ($\chi^2=11$), although
the GHRS equivalent widths agree with the prediction of full ionization.  
We note that interstellar and photospheric absorption lines, if present,
provide additional absorption components that are not expected to show
strong orbital variation.  If these components are present, this would
naturally explain why the fractional variation in equivalent width that we
see is less than that predicted by $q\ll1$.  For example, for
\ion{N}{5}, $\bar{W_\lambda}=2.4$\AA, while $W_\lambda$ due to 
photospheric absorption was found with the GHRS to be $\approx0.8$\AA.
Without consideration of the photospheric
absorption, the observed variation of 1.6\AA\  corresponds to
$\Delta=67$\%.  However, this corresponds to a 100\%\ variation in the
equivalent width of absorption due to the wind alone, excluding the
photospheric absorption.

It is likely that Figure~4 underestimates the per cent variation of the
P~Cygni absorption for the \ion{C}{4} and \ion{Si}{4} lines, as X-ray
should be {\it more} effective in removing these ions from the wind than
removing \ion{N}{5} from the wind.  A systematic uncertainty in choosing
the continuum level will have a greater effect on the fractional variation
in the equivalent width of the weaker lines, such as \ion{Si}{4}.

In the next section, we will make the simplifying assumption that the
entire region outside the conical shadow of the normal star is ionized by
the X-ray source, and from this derive the radial velocity law.

\subsection{The radial velocity law from the maximum
velocity of absorption\label{sec:vlaw}}

The spectra obtained with the GHRS were read out at 0.5 second intervals,
using its RAPID mode (Paper~I).  The P~Cygni absorption profile was
variabile on time scales shorter than the HST orbit.
We interpret this as the result of the X-ray shadow of the primary
crossing the cylinder subtending the face of the primary, where the
blue-shifted absorption is produced.  We attempt to infer the radial
velocity law in the wind from this rapid P~Cygni variation and a few
simple assumptions. 

We assume here that X-ray photoionization removes \ion{N}{5} and
\ion{C}{4} from the
entire wind outside of the X-ray shadow of the primary.  This assumption
is easily shown to be reasonable, based not only on the IUE and HST
equivalent width variations (\S\ref{sec:IUE}) but also on the 
expected value of $q$.  Given a mass-loss rate of 
10$^{-6} \Msun$~yr$^{-1}$, a velocity at the X-ray source of
600~km~s$^{-1}$, and an ionization parameter $\log \xi=2.5$ at which 
there should be little \ion{N}{5} present (Kallman \&\ McCray 1982),
we find $q=0.16$.

As Buff \&\ McCray (1974) pointed out, as an X-ray source orbits a star
with a strong stellar wind, the absorbing column to the X-ray source
should vary in a predictable way that depends on the wind structure and
the orbital inclination.  If the wind is not photoionized, then the
absorbing column increases near eclipse.
In a companion paper (Boroson et al. 1999) we will examine the ionization
in the wind from the variation in the absorbing column as seen with ASCA.  
Preliminary analysis supports $q\ll1$.

We now use the geometry shown in Figure~5.  The primary star
is assumed to be spherical with a radius $R$.  The neutron star orbits at 
a radius $D$.  The angle between the ray connecting the neutron 
star to the center of the primary and a tangent to the primary's surface is 
given by $\cos \theta=R/D$.  Then we define three relevant distances:
\begin{eqnarray}
D_2 & \equiv & \frac{R}{\cos(\pi-\theta-2\pi\phi)}\\
D_3 & \equiv & D_2+\frac{R D_2}{D \sin 2 \pi \phi}+\frac{R}{\tan 2 \pi 
\phi},\\ 
D_4 & \equiv & (D_3^2+R^2)^{1/2}.
\end{eqnarray}

We wish to compute, given $\phi$, the radius $r(\phi)$ in the wind
responsible for the maximum observed velocity of absorption.  Given
$(r(\phi),v(\phi))$ derived from closely spaced $\phi$ when the shadow of
the primary is moving across the line of sight, we can compare with the
theoretical expectation of the wind acceleration velocity law.  (Note that
this assumes a spherically symmetric wind.) 

Given a monotonic velocity law, in most situations the maximum absorption
velocity will be observed to be $V(\phi)$ corresponding to a wind radial
velocity $V_{\rm max}$ (Figure~5), at a distance of $D_4$ from the center
of the primary.
This is the furthest
radial distance at which the wind can absorb stellar continuum and at
which the X-rays from the neutron star are shadowed by the limb of the
primary.  Note that the wind at this point is moving obliquely so that 
$V(\phi)=V_{\rm max} D_3/D_4$.  As
$\phi$ increases from 0.1 to 0.5, the wind at $V_{\rm max}$ will move with
increasing obliqueness to the line of sight.  As a result, even though
this point has the greatest radial distance of any point in the wind that
can absorb \ion{N}{5}, it still may not cause the absorption with the
highest
possible {\em line of sight} velocity, even in a monotonic wind.  As a
first check to see if the obliquity of the velocity vector at $V_{\rm
max}$ is important, we can compare $V(\phi)$ with $V_{\rm line-of-sight}$,
which can be
calculated from $D_2$, assuming a radial velocity law. 

We assume that $D/R=1.77$, which can be calculated from the mass ratio and
the assumption that the primary fills its Roche lobe.
(Note that we use the separation between the
centers of mass of the two stars, and
not the distance of the NS from the CM of the system, which is about
10\%\ smaller.) 

We determine the maximum velocity of absorption by fitting the
high-velocity edge of the absorption profile with a broken line
(Figure~6). The line
is horizontal for the continuum blue-ward of the maximum velocity of
absorption and has a nonzero slope, allowed to vary as a free parameter,
at velocities less than the maximum.  When the doublets do not overlap
($\phi>0.19$), we are able to obtain two maximum velocities at each phase,
one for each doublet component.  For the red doublet component, we do not
fit a broken line in order to determine the maximum absorption velocity,
as the profile is never horizontal.  Instead, we find the maximum
absorption velocity in the
red doublet component by determining where the flux rises to the
continuum level (which is determined by the broken-line fit to the blue
doublet component.)  The results from the red and blue doublet components
do not show a systematic difference.  Thus averaging maximum
absorption velocities determined in each line allows us to
reduce the errors in our velocity measurements.

We can apply a further refinement to the method.  The observed P~Cygni
trough is actually a sum of the true absorption profile and blue-shifted
P~Cygni emission.  The observed maximum absorption velocity may be shifted
by the addition of the emission.  Although we can not observe the
blue-shifted emission profile directly (it is masked by the absorption
whose profile we are trying to uncover) we can approximate it using the
reflected red-shifted profile at a complementary orbital phase, assuming
spherical symmetry and the Hatchett-McCray effect.  The emission flux $F$
at velocity $v$ and at phase $\phi$ is given by \begin{equation}
\label{eq:emitcorrect} F(v,\phi)\approx F(-v,0.5-\phi).  \end{equation}
For example, at $\phi=0.1$, the red-shifted emission at high velocities
will be diminished as a result of photoionization; the blue-shifted
emission at $\phi=0.4$ will be similarly diminished.  (This symmetry is
broken because the primary shadows a portion of the red-shifted emission
but none of the blue-shifted emission.  However, we can safely neglect
this effect because the blue-shifted emission in front of the stellar disk
can not have a higher velocity than the blue-shifted absorption.)

The zero-velocity for the reflection of the red-shifted
emission is not completely constrained.  We allow two choices: 1) that the
systemic velocity of LMC~X-4 is identical to that of the LMC, and 2) that
the systemic velocity of LMC~X-4 is 150~km~s$^{-1}$ so that the turnover
between red-shifted emission and blue-shifted absorption occurs near a
velocity of~0.

We give the measured values of $(r(\phi),v(\phi))$ in Table~1.  We
consider that a reasonable $1\sigma$ error for the velocity measurements
is 50~km~s$^{-1}$.

As an independent check on our measurements of wind velocity, we examine
the maximum velocities at which the absorption profiles vary between two
orbital phases.  This is the only method we can apply to the
\ion{C}{4}$\lambda1548$ lines, in which the P~Cygni absorption can not be
separated from the strong interstellar and photospheric lines by
examination of the spectrum 
alone.  The best-fit maximum velocity of the \ion{C}{4} line is
at $910\pm50$ km~s$^{-1}$.  We note that we did not observe any
red-shifted P~Cygni emission in \ion{C}{4}, so were unable to compensate
for
the variability in the blue-shifted emission.

The maximum velocity of the change in \ion{N}{5} absorption
between $\phi=0.111$ and $\phi=0.209$ is $1130\pm50$ km~s$^{-1}$, 
consistent with the 
measurements in Table~1.  Between $\phi=0.209$ and 
$\phi=0.312$, the peak velocity of absorption variations is $560\pm50$
km~s$^{-1}$,
consistent with the previous measurements (depending on the model of the
blue-shifted emission that is subtracted). However,
between $\phi=0.312$ and $\phi=0.410$, the maximum velocity of the
absorption variations is $160$~km~s$^{-1}$.
This is lower than the velocity inferred from
the line profiles alone ($\approx320$ km~s$^{-1}$ at $\phi\approx0.312$).

As discussed in \S\ref{sec:photo}, the blueshifted absorption may have a
component which is produced near the stellar photosphere, rather than in
the wind.  In this case, the velocity width of the line may be produced by
the rotation of the star, rather than the wind expansion.
We suggest that the absorption line for $\phi>0.25$ is likely to contain a
large photospheric component.  Thus we do not fit to the maximum
absorption velocity at these phases.

We fit our measurements of $(r(\phi),v(\phi))$ to the analytic expression
\begin{equation}
\label{eq:windvlaw}
v=v_{\infty} (1-1/x)^\beta+v_0 \end{equation} where $v_0$ is the systemic
velocity of LMC~X-4, and $x=r/R$.  The best-fit values and 
formal errors are
presented in Table~2, for 3~possibilities of organizing the data,
depending on whether we subtract our model for the blue-shifted emission
from the spectra and whether the velocity of the reflection point is 0
or 150~km~s$^{-1}$.  We find the best fit using a downhill simplex method
(Nelder \&\ Mead, 1965) and estimate errors using a Monte-Carlo bootstrap
method.


We plot $(r(\phi),v(\phi))$ (measured from both
doublet components) in Figure~7, along with the analytic fits.
There is no evidence that the wind is nonmonotonic.
All the results indicate that the wind expands more slowly than expected
from the radiatively accelerated solution (which can be well-approximated
with $\beta=0.8$;
Pauldrach, Puls, \&\ Kudritzki 1986).
While there is a large range of $\beta$ that can
fit the 3 arrangements of observing $(r(\phi),v(\phi))$ ($1.4<\beta<2.4$), 
the choice of $\beta=1.39\pm0.14$ provides the lowest $\chi^2$ and the
most reasonable systemic velocity. (However, note that turbulence in the
wind could add a constant
velocity to the absorption edge and could cause the data to be well-fit by
a nonphysical value of the systemic velocity.)  The models are also
consistent with the measurements of the wind veocity from differences of
successive spectra of \ion{C}{4} at $\phi=0.15$ and \ion{N}{5} at
$\phi=0.3$.

While a coincident stellar absorption feature at $\lambda\approx1249$\AA\
decreases the red-shifted emission for velocities $v\gtrsim1000$~\kms,
compensating for this feature in our subtraction of the blue-shifted
emission would only {\it increase} slightly the value of $\beta$ that we
measure.

Our fits for the stellar wind acceleration parameter $\beta$ are not
sensitive to our choice of $D=1.77$; using $D=1.70$ and $D=1.84$ we find
$\beta$ within $1\sigma$ of values in the first two rows of Table~2,
but $\beta=2.00$ instead of $2.37\pm0.11$ for the third row.  

The effects of microturbulence in the wind on the measured maximum wind
velocity should not alter the values we have found for $\beta$.  For
example, if the microturbulence velocity is constant, then our fits would
give a different value of $v_0$ but the same $\beta$.  If the
microturbulence velocity is a constant fraction of the wind velocity, then
as a rough approximation, all measured values of $v(\phi)$ will be
multiplied by a constant, leading us to fit $(r(\phi),v(\phi))$ with a
different value of $v_\infty$ but the same value of $\beta$.

In this section we have demonstrated a new method of probing the velocity
field of a stellar wind.  We have assumed only standard parameters of the
stellar radius and separation, that the wind is spherically symmetric and
that the X-ray source completely ionizes the wind.  One drawback of the
method is that only a very small portion of the available data is used,
that is, only the velocity of the edge of the absorption.  In the next
section, we adapt the method to fit the entire absorption profile. 

\subsection{The Radial Velocity Law from the Absorption Line Profile
\label{sec:wholeprof}}

We now attempt to fit the entire absorption profile to determine the wind
radial velocity law.  The geometry of this method is shown in Figure~8. 
The X-ray shadow of the primary is conical.  At distances $r\gg R$ from
the center of the primary, the wind within the cylindar subtending the
stellar surface has only a small oblique velocity (assuming a radial
velocity field).  Thus the surfaces of constant projected velocity, far
from the primary, are planes perpendicular to the line of sight.  The
intersection between the cone of shadow and the planar isovelocity surface
at a given observed velocity will most often be a parabola.  This parabola
bounds those portions of the stellar disk that lie behind unilluminated
gas at the observed velocity, and thus are absorbed by the \ion{N}{5} in
the
wind, and those portions that do not. 

If we could obtain an image of the stellar disk of the primary in LMC~X-4
with a narrow-band filter centered at a wavelength corresponding to a
Doppler shift at the given velocity, we would see a crescent-shape bounded
by the star's disk and the parabola.  As the wavelength of the filter
approached the terminal velocity of the wind, the intersection of the
plane and cone would bound a smaller and smaller region of the stellar
surface, and would thus determine the shape of the edge of the P~Cygni
profile. 

In general, the isovelocity surfaces will not be planes, and we will need
to solve for them numerically.  We can no longer simply read off the
velocity corresponding to each radius from the absorption profile, as we
did in \S\ref{sec:vlaw}. Instead, we need to assume a velocity law,
find the isovelocity surfaces, find their intersection with the conical
shadow, and then compute the absorption profile and compare with the
observed profile.  

To use the observed absorption profile, we must correct for the
blue-shifted emission by subtracting the red-shifted emission at a
complementary phase.  Now that we consider the line profile and not
merely the edge of the absorption, we must 
recognize the asymmetry between the red and blue-shifted
emission profiles at $\phi$ and $0.5-\phi$
which is introduced by the primary's shadow of the red-shifted emission.
We correct for this
by assuming that the shadowed red-shifted
emission at the complementary phase is given by 
\begin{equation} F_{\rm
shadow} (v,\phi) = -\frac{1}{2}\left(1-\frac{\sqrt{x^2-1}}{x}\right)F_{\rm
abs} (-v,0.5-\phi)  
\end{equation} 
where $x$ is the radial distance in the
wind, normalized to the stellar radius, $F_{\rm shadow}$ is the line
profile of the shadowed part of the red-shifted emission, and $F_{\rm
abs}$ is the intrinsic absorption line profile.  For this approximation,
we have noted that the shadowed red-shifted emission at $\phi$ corresponds
to the blue-shifted absorption at $0.5-\phi$, with the dilution factor
needed to compare the absorption with the emission.

We also need to assume an optical depth in the wind.  As a first attempt,
we assume that the wind is entirely black to the continuum in the region
of \ion{N}{5}.  We neglect limb and gravity darkening of the primary, and
its
nonspherical shape.  We correct for the photospheric absorption by
subtracting the profile at $\phi=0.4$, when we assume the absorption is
entirely photospheric.

A fit to individual short subexposures, as in \S\ref{sec:vlaw} would not
be time-efficient.  Instead, we fit to the blue doublet component
absorption profiles at $\phi=0.1,0.2,0.3$.  The fit is poor, with
$\chi^2_\nu=11$ with 253 degrees of freedom, $v_\infty=1780\pm10$ \kms,
$v_0=110\pm10$ \kms, and $\beta=4.25\pm0.08$, which we do not consider a
reasonable value.  Allowing a uniform optical depth $\tau$ in the wind
gives a better fit, with $\chi^2=2.95$, $v_\infty=1330\pm30$ \kms,
$v_0=72\pm24$ \kms, $\beta=1.57\pm0.04$, and $\tau=0.97\pm0.06$.  We show
the results of this fit in Figure~9.  This high value of $\beta$ is
similar to those found in \S\ref{sec:vlaw}.

In the following section, we will
investigate detailed numeric models of the wind, relaxing our assumption
that the X-rays ionize the entire wind outside the X-ray shadow of the
normal star.
We do not remove
the restriction that the wind velocities are 
spherically symmetric, as this would allow
more than one point along the line of sight to be at the same projected
velocity, which would require us to use more sophisticated but 
time-consuming techniques, such as Monte-Carlo simulation of P~Cygni 
profiles.

\section{Escape Probability Models\label{sec:numeric}}

\subsection{Methods and basic models\label{sec:basic}}

To model the P~Cygni line variations, we apply and extend the method of 
McCray et al. (1984).  We compute the line profiles using the Sobolev 
approximation and the escape probability method
(Castor 1970; Castor \&\ Lamers 1979), using a wind 
velocity law given by Equation~\ref{eq:windvlaw} with $\beta=1$.
The optical depth in the undisturbed 
wind obeys 
\begin{equation}
\label{eq:tau}
\tau=T(1+\gamma)r^{-\gamma}
\end{equation}
where $T$ and $\gamma$ are empirical parameters (Castor \&\ Lamers 1979).

The optical depth of gas illuminated by the X-ray source is given by the
Sobolev approximation, 
\begin{equation} 
\label{eq:sobolev}
\tau=(\pi e^2/m c)f \lambda_0 n_{\rm i,j} (dv/dr)^{-1}, 
\end{equation} 
where $e$ is the electron charge,
$m$ the electron mass, $\lambda_0$ the rest wavelength of the transition,
$f$ the oscillator strength, and $n_{i,j}$ the particle density of
element $i$ in ionization state $j$.  From mass conservation, $n_{i,j}$ is
given by 
\begin{equation}
\label{eq:density}
 n_{\rm i,j} = \frac{\dot{M}}{4 \pi r^2 v} m_{\rm
H}^{-1} a_{\rm i} g_{\rm i,j},
\end{equation}
where $g_{i,j}$ is the fraction of element $i$ in ionization state $j$,
and $a_{\rm i}$ is the fractional abundance of element $i$. 
Given a value of $\xi$ and the shape of the X-ray spectrum, the ion fraction
$g_{i,j}$
(and thus $n_{\rm i,j}$) can be computed.  For this we used the XSTAR code 
(Kallman \&\ Krolik, based on Kallman \&\ McCray 
1982), which requires that the X-ray luminosity $L_{\rm x}$ be integrated 
over the energy range 1--1000~Ry (13.6 eV--13.6 keV).
The effectiveness of X-ray photoionization 
depends on the X-ray spectrum; we have used a broken power-law spectrum,
\begin{eqnarray}
F(E) & = & K E^{0.37}\qquad\mbox{for }E<25\mbox{ keV}\\
     & = & K E^{-2}\qquad\mbox{for }E>25\mbox{ keV}, \end{eqnarray} where
$F(E)$ is the energy spectrum. 
The power-law index for $E<25$~keV was determined from simultaneous ASCA
observations of the 0.4--10 keV spectrum (Paper~I), while the high-energy 
cutoff has been observed at other times
using GRANAT (Sunyaev et al. 1991).  Adding a soft blackbody that
contributes $\sim10$\%\ to the total flux (as seen in the X-ray spectrum
with ASCA) has little effect on the ion fractions.  We can treat the
combined effect of the background ionization (due to the photospheric EUV
flux or X-ray emission from shocks in the wind for example) and the X-ray
ionization by 
\begin{equation} \label{eq:background1} g_{i,j,\rm
tot}=(1/g_{i,j,\rm X} +1/g_{i,j,\rm back} - 1)^{-1}, \end{equation} 
in the
case in which the wind in the absence of X-ray illumination is at least as
highly ionized as the ion which causes the P~Cygni line.  
If this is not
the case, then we have 
\begin{equation} \label{eq:background2}
1/g_{i,j,\rm tot}-1=(1/g_{i,j-1,\rm X}+1/g_{i,j-1,\rm back} - 2)^{-1}
\end{equation} 
In Equations~\ref{eq:background1} and~\ref{eq:background2},
$g_{i,j,\rm X}$ and $g_{i,j,\rm back}$ are, respectively, the ion
fractions with X-ray illumination (computed using XSTAR) and without
(computed from equations~\ref{eq:tau}, \ref{eq:sobolev}, and 
\ref{eq:density}).     

We assume that all N in the unilluminated wind is ionized to at least
\ion{N}{5} and apply Equation~\ref{eq:background1}.  The models of
$\tau$Sco ($T\approx$30,000) by MacFarlane, Cohen, \&\ Wang (1994) predict
that the dominant ionization stage of N in the wind is \ion{N}{6} for wind
velocities $v>0.2v_\infty$.  In these models, the wind is ionized by
X-rays, presumably caused by shocks in the wind, that have been observed
by ROSAT. However, MacFarlane et al. found that their model of $\tau$Sco
did not match UV observations, and that X-rays were less effective in
ionizing the denser winds of hotter stars (consistent with the results of
Pauldrach et al. 1994). We also tested models in which \ion{N}{4} was the
dominant ionization stage of the unilluminated wind, and found that the
best-fit parameter values were within the errors of the model we present
here.  The choice of Equation~\ref{eq:background1} or
Equation~\ref{eq:background2} has a small effect on our results, as it
affects only that portion of the wind that is illuminated by direct X-rays
from the neutron star, yet is ionized so slightly that the background
ionization rate is important.

We treat the overlap of doublet components using the 
approximation of Castor \&\ Lamers (1979).  The optical depths in the
photospheric absorption lines are assumed to have Gaussian profiles.
The red and blue doublet components of the photospheric lines have 
optical depths at 
line center of $\tau_{\rm b}$ and $\tau_{\rm r}=0.5 \tau_{\rm b}$ (in 
the lines we consider, the oscillator strengths are in a 2:1 ratio).

We do not account for limb and gravity darkening of the companion or its
nonspherical shape.  However, we normalize the absorbed continuum (which
arises in the visible face of the star) to the continuum level observed by
the GHRS at each particular orbital phase, while we normalize the
scattered emission (which can originate anywhere on the companion) to the
average continuum level observed with the GHRS.  

The free parameters of the fit are $T$, $\gamma$, $v_{\infty}$, $v_0$ (the
systemic Doppler shift of LMC~X-4), $L_{38}/\dot{M_{-6}}$ (the ratio of
X-ray luminosity in 10$^{38}$ erg~s$^{-1}$ to mass-loss rate in 10$^{-6} 
\Msun$~yr$^{-1}$), $\dot{M_{-6}}$,
and $\tau_{\rm b}$.  We employ a 
downhill simplex method (Nelder \&\ Mead, 1965) 
to find the minimum $\chi^2$ and then refine the solution and find
errors in the parameters using the Levenberg-Marquart method (Marquardt,
1963).  For all velocities, we set the error to at least 35~km~s$^{-1}$,
which is the resolution limit of the GHRS G160M grating. We exclude from
the fit the central 1\AA\ region around the rest wavelength of each
transition, which is poorly modeled by the escape probability method. 

In Table~3 we show the best-fit parameters and we display the fits to the
line profiles in Figure~10.

The fits determine $L_{38}/\dot{M_{-6}}=0.26\pm0.01$ because the ion
fractions in the wind (and thus the optical depths) depend sensitively on
this ratio.  The fits also determine $\dot{M_{-6}} a_{\rm N}/a_{\rm N,
LMC}=3.2\pm2.3$, where $a_{\rm N, LMC}=1.8\times10^{-5}$ is the Nitrogen
abundance in the LMC, assumed to be 20\%\ of the cosmic value.  The error
determined for this parameter shows that it is not effectively constrained
by the fit;  this is because it only affects linearly the optical depth in
the portion of the wind outside of the X-ray shadow. 

In general, the model successfully matchess the change in the P~Cygni 
profiles between successive orbital phases (Figure~10, right-hand panel).
However, the predicted change in the \ion{N}{5} profiles between
$\phi=0.31$ 
and $\phi=0.41$ is blue-shifted from the observed change.  This is 
further evidence that the wind expands more slowly at small radii than 
expected from a $\beta=1$ velocity law.  

\subsection{Wind Inhomogeneities\label{sec:inhom}}

We now allow the wind density to be inhomogeneous by adding two new free
parameters, another mass-loss rate $Z \dot{M}$, where $Z\gg1$, and a
covering fraction $f$.  In the absence of X-ray photoionization, the
optical depth in the two wind components are $\tau_1=\tau$ (where $\tau$
is given by Equation~\ref{eq:tau}) and $\tau_2=Z \tau$.  The ionization
fractions are computed for a given point in the wind that has a density
determined from $\dot{M}$ and $Z \dot{M}$.  For the blue-shifted
absorption, 
a fraction $f$ of sightlines to the stellar surface encounters material
with the higher density.  This material has a much higher optical depth,
as not only is there more of this material, but it also contains a higher
fraction of \ion{N}{5}. 

Applying a downhill simplex $\chi^2$ minimization algorithm gives an
improved $\chi^2_\nu=4.01$.  The fit is improved for two reasons.  First,
the fit of the homogeneous wind model over-predicted the effects of X-ray
ionization on the red-shifted emission (Figure~10).  The inhomogenous wind
model allows more \ion{N}{5} to survive in the X-ray illuminated wind, so
that more of the red-shifted emission persists.  Second, the homogeneous
wind model predicts too much absorption in the 1238\AA\ line and not
enough absorption in the 1242\AA\ line at $\phi=0.12$.  When an
inhomogeneous wind is allowed, the strength of the absorption in the two
components can be more nearly equal, as absorption from the dense wind is
saturated.  We compare the inhomogeneous wind models with the GHRS
observations in Figure~11. 

We find a ``covering fraction'' $f=0.33\pm0.01$; this is the probability
that a ray from the primary encounters denser material with the mass-loss
rate $Z\dot{M}$.  Our best-fit values are $Z=700\pm200$ and
$L_{38}/\dot{M_{-6}}=160\pm40$.  If we assume that the covering fraction
is approximately the volume filling factor, then the average mass-loss
rate of the wind, $\bar{\dot{M}}=(1-f)\dot{M}+f Z \dot{M}$, then leads to
$L_{38}/\bar{\dot{M{-6}}}=0.7\pm0.4$. 

We note that the best-fit value for the photospheric optical depth in the
blue doublet component, $\tau_{\rm b}=0.62\pm0.02$ is higher than the
values obtained for similar stars by fits to IUE data (Groenewegen \&\
Lamers 1989).

\subsection{Alternate Velocity Laws\label{sec:beta}}

As discussed in \S\ref{sec:vlaw}, the change in the maximum velocity of
the P~Cygni line absorption over orbital phase allows us to infer the
radial velocity law in the wind, assuming only that the X-ray source
ionizes the entire wind outside of the shadow of the normal star.  In
\S\ref{sec:basic}, we modelled the P~Cygni line variations allowing the
X-ray photoionization of the wind to vary but assuming a standard
$\beta=1$ radial velocity law.  If we allow a slowly accelerating wind
with $\beta=1.6$ we can still obtain a good match with the observed line
profiles, with $\chi^2=4.91$.  In this case we find
$v_\infty=1560\pm35$~\kms,
$L_{38}/\dot{M}_{-6}=0.23$
Allowing an inhomogeneous wind improves the fit
to $\chi^2=4.59$.  We conclude that while the line profiles can be
adequately fit with a $\beta=1$ velocity law,
a more slowly accelerating wind,
inferred from simpler models, is still consistent with the data. 
Velocity laws with
$\beta>2$ (as in some of the fits in \S\ref{sec:vlaw}) generally predict
steeply peaked red-shifted emission (Castor \&\ Lamers 1979); however this
is difficult to rule out from the data, as
photospheric absorption is probably present at low velocities.

\subsection{X-ray Shadow of the Accretion Disk and Gas Stream on the
Wind\label{sec:diskjet}}

The accretion disk or gas stream from the L1 point could presumably shadow
regions of the stellar wind from X-ray illumination.  We have made models
to find signatures of these effects in the P~Cygni line profiles.

LMC~X-4 shows a $\approx30$~day variation in its X-ray flux (Lang et al. 
1981), usually attributed to shadowing by an accretion disk that we
observe at varying orientations.  To allow the disk to shadow regions of
the wind from the X-rays, we make the simple assumption that it is flat
but inclined to the orbital plane.  Howarth \&\ Wilson (1983) used a
similar geometry to model the optical variability of the Hercules~X-1
system, which shows a similar 35~day X-ray period.  We allow as free
parameters the angular half-thickness of the disk, its tilt from the
orbital plane, and its ``precessional'' phase within the 30-day cycle. 
Fixing the 30-day phase to be the at the X-ray maximum (as suggested by
observations with the All Sky Monitor on the Rossi X-ray Timing Explorer,
Paper~I), we perform a $\chi^2$ fit and find a disk half-thickness of
$6.9\pm0.3\arcdeg$ and
a disk tilt of $31\pm2\arcdeg$ from the orbital plane, for a reduced
$\chi^2=4.62$.

To simulate the shadow caused by the gas stream from the L1 point, we have
assumed that the stream emerges at an angle of $22.5\arcdeg$ from the line
to the neutron star (Lubow \&\ Shu 1975).  The stream blocks the X-rays
from a uniform half-thickness $\rho$ in elevation from the orbital plane
(as seen from the X-ray source) and at angles $<\eta$ from the line to the
L1 point along the orbital plane; $\rho$ and $\eta$ are free parameters of
the fit.  The fit gives $\chi^2=4.04$, with $\rho=33\pm2\arcdeg$ and
$\eta=90\pm3\arcdeg$.  The values for the other free parameters similar to
those found in \S\ref{sec:basic}. 

The reason that the gas stream shadow improves the P~Cygni profile fits is
that at $\phi=0.1-0.4$, the stream shadows the receding wind, increasing
the red-shifted emission, which our basic model of \S\ref{sec:basic}
underpredicts.  Allowing an inhomogeneous wind as in \S\ref{sec:inhom}
improves the fit for the same reason.  It is possible that both effects
are combined, and future observations will be needed to disentagle them.

\subsection{\ion{C}{4} Lines\label{sec:c4}}

The \CIV lines did not show as much variability with the GHRS as the
\ion{N}{5}
lines, although we did not observe the lines before $\phi=0.16$.  Most of
the absorption in these lines is probably either interstellar or
photospheric (Vrtilek et al. 1997).  Nevertheless, as a check on our
models of the \ion{N}{5} lines, we also performed fits to the \CIV lines.
We fix the parameters found in \S\ref{sec:basic},
except for $T$, $\gamma$, and
$\tau_{\rm b}$, which are free parameters defining the optical depth of
\ion{C}{4} in the wind and photosphere.  We added fixed Gaussian
interstellar
\CIV absorption lines to the profiles.

The result of our model, shown in Figure~12, does not match the line
profiles in detail ($\chi^2=11.3$).  It is possible that the fit is poor
because there are photospheric lines in the vicinity of 1550\AA\ due to
ions other than \ion{C}{4}.  A systemic velocity closer to 0 might bring
the photospheric lines into closer agreement with the data.  In both the
observed profile and in the model, red-shifted emission is not prominent.
The fitted value of $\gamma=7.9\pm0.8$ implies that \ion{C}{4} in the
undisturbed wind is confined to $x<2$, so that much of the red-shifted
\ion{C}{4} emission could be occulted by the primary.  If we fit the line
profiles with models with higher values of $\beta$, then $\gamma$ is
reduced, but our fits never gave $\gamma<4.5$ for $\beta<2.4$ (the upper
limit found in \S\ref{sec:vlaw}).


We compare our values of $\gamma$ with the optical depth law found for
\CIV by
Groenewegen \&\ Lamers (1989) by fitting the \CIV P~Cygni lines of similar
stars observed with IUE.  They parameterize the optical depth in the
stellar wind by \begin{equation} \tau\sim (v/v_\infty)^{\alpha_1}
[1-(v/v_\infty)^{1/\beta}]^{\alpha_2} \end{equation} where
$\alpha_2=\gamma$ for
$\alpha_1=0$ and $\beta=1$.  For HD~36861 (spectral
type O8\,III) Groenewegen \&\ Lamers find $\alpha_1=1.5\pm0.2$ and
$\alpha_2=3.1\pm0.5$, which is inconsistent with $\gamma>3.1$.  For
HD~101413, they find $\alpha_1=-0.8\pm0.3$ and $\alpha_2=1.6\pm0.4$.  
These values imply a sharply peaked concentration of C\,IV towards the
stellar surface, but do not fit our data as well as $\gamma\gae5$.

We have no compelling alternate explanation for our high value of
$\gamma$.  However, the \CIV lines in LMC~X-4 that we observed were weak
and were dominated by photospheric absorption.  Further observations at
phases when \CIV is more prominent are needed to test our conclusion that
this ion is concentrated near the stellar surface.

\subsection{Fits using the SEI method}

It has been shown (Lamers et al. 1987) that inaccuracies resulting from
the escape probability method can be reduced by computing the radiative
transfer integral exactly along the line of sight while continuing to use
the source function given by the Sobolev approximation.  This is called
the SEI (Sobolev with Exact Integration) method.  The method can be used
to simulate the effects of local turbulence within the wind, and using the
method of Olson (1982), the effects of overlapping doublets can be
computed precisely.

We have implemented a program that uses the SEI method to predict the
P~Cygni line profiles from a wind ionized by an embedded X-ray source. 
The details of the wind ionization are identical to those given by
Equations~\ref{eq:sobolev} through \ref{eq:background2}.  We find that in
our analysis of LMC~X-4, the approximation we have made for the doublet
overlap gives results very close to those given by Olson's method and the
SEI method.  This may result from the small amount of the doublet overlap
in the \NV lines, given a separation of 960~km~s$^{-1}$ between the
doublet components and a wind terminal velocity of 1350~km~s$^{-1}$. 
Using a turbulent velocity $v_{\rm t}$ that is constant throughout the
wind, we find that the line profiles are best fit with $v_{\rm
t}<200$~km~s$^{-1}$.

The SEI method also confirms the results of the escape probability model
for the \CIV lines, as reported in \S\ref{sec:c4}.  Although the
separation of the \CIV lines is only 500~\kms, we suggest that the doublet
overlap was not severe during our observations because much of the wind at
high velocity had already been ionized, and \ion{C}{4} is concentrated
near the primary's surface, at low wind velocities.

\section{Discussion}

We have demonstrated a new method for inferring the radial velocity
profile in a stellar wind.  The wind radial velocity law for LMC~X-4 can
be fit with
$\beta\approx1.4-1.6$, and $\beta<2.5$.  This differs from the expected
value of $\beta=0.8$, but the wind in the LMC~X-4
system is likely to deviate from that of an isolated O star.  Assumptions
of the method, that the wind is spherically symmetric, for example, will
need to be revised in light of further observations that cover the entire
orbital period.

Unfortunately, we did not observe the UV spectrum during the X-ray eclipse
when the X-ray ionization of the blueshifted absorption should be minimal,
and so our inference of the wind terminal velocity is model-dependent. 
Observations with STIS and the Far Ultraviolet Spectroscopic Explorer
(FUSE)  during eclipse may distinguish between our alternate models.  The
wind terminal velocity we have inferred from our $\beta=1$ escape
probability method, $v_\infty=1350\pm35$~\kms, is lower than the terminal
velocities measured for similar stars by Lamers, Snow, \&\ Lindholm
(1995).  Their stars \# 33,37,38,39, and 44 are all near the 35,000\,K
temperature of LMC~X-4, but have $v_\infty=1500-2200$~\kms.  However,
stars in the LMC are known to have terminal velocities $\approx20$\%\
lower than their galactic counterparts (Garmany \&\ Conti 1985).

We note from the IUE data shown in Figure~4 that the P~Cygni line
absorption at orbital phases $\phi>0.5$ is greater than that at
$\phi<0.5$.  X-ray absorption dips are frequently seen at $\phi\sim0.8$,
possibly indicative of dense gas in a trailing gas stream or accretion
wake.  However, it may be difficult for models to reproduce an accretion
wake that occults a large fraction of the primary at orbital phases as
late as 0.9.  Another possible explanation is that a photoionization wake
is present (Fransson \&\ Fabian 1980, Blondin et al.  1991).  Such a wake
results when the expanding wind that has not been exposed to X-rays
encounters slower photoionized gas.  While a strong photoionization
wake is expected for $q\ll1$, the presence of Roche lobe overflow and a
photoionization wake have been found to be mutually exclusive in some
simulations (Blondin et al. 1991).  A further possibility is that at
$\phi\sim0.6-0.9$ the wind in the cylindar subtending the primary's disk
is shielded from ionization by the trailing gas stream.  A simple
prediction of this scenario is that there should be more flux in the
red-shifted emission at $\phi=0.25$ than at $\phi=0.75$.  If the primary
star does indeed rotate at half the corotation velocity (Hutchings et al. 
1978) then the gas stream may trail the neutron star by more than the
22.5$\arcdeg$ predicted by Lubow \&\ Shu (1975). 

We have fit the changing P~Cygni line profiles using an escape probability
method, assuming a radial, spherically symmetric stellar wind.  However,
we note that the Hatchett-McCray effect is most sensitive to the region
between the $q=1$ surface and the conical shadow of the primary.  Outside
of this region, the wind should be entirely ionized for likely values of
$\dot{M}$ and $L_{\rm x}$.  The region of the wind that our method is
sensitive to is illuminated by X-rays, and its dynamics may be affected as
a result (Blondin et al.  1990).  Structure associated with the distortion
of the primary, such as the gas stream and a wind-compressed disk
(Bjorkman 1994), may be also be important in this region.

We have inferred $L_{38}/\dot{M}_{-6}=0.26\pm0.01$ from our escape
probability P~Cygni line models.  Our simultaneous ASCA observations found
a 2--10~keV flux of $2.9\times10^{-10}$~erg~s$^{-1}$~cm$^{-2}$, which
given a distance to the LMC of 50~kpc implies $L_{38}=0.9$ and
$\dot{M}_{-6}\approx3$ (our inhomogeneous wind model of \S\ref{sec:inhom}
would imply $\dot{M}_{-6}\approx1$).  There is some uncertainty in
applying $L_{38}=0.9$, as the isotropic X-ray spectrum and luminosity may
differ from that observed by ASCA.  Nevertheless, the derived $\dot{M}$ is
an order of magnitude larger than that reported by Woo et al. (1995), who
used the strength of scattered X-rays to find
$\dot{M}/v_\infty=10^{-10}\Msun$~yr$^{-1}$~(\kms)$^{-1}$, implying for our
best-fit value of $v_\infty=1350\pm35$ km~s$^{-1}$ that
$\dot{M}_{-6}=0.14$.  Our result for $\dot{M}$ is on the order of the
single-scattering limit given by the momentum transfer from the radiation
pressure of the primary;  $\dot{M}<L_{\rm opt}/v_\infty c$, where the
optical luminosity of the primary $L_{\rm
opt}\sim5\times10^{38}$~erg~s$^{-1}$ (Heemskerk \&\ van Paradijs 1989) 
implies that $\dot{M}<2\times10^{-6}\Msun$~yr$^{-1}$.  However, if the
wind is accelerated by scattering of overlapping lines, one can have
$\dot{M}= 2-5 L_{\rm opt}/v_\infty c$ (Friend \&\ Castor 1982). 

Combining our value of $\dot{M}_{-6}\approx3$ with our measured wind
velocity near the orbit of the neutron star (400~\kms) and the orbital
velocity of the neutron star (440~\kms), we find that gravitational
capture of the stellar wind (Bondi \&\ Hoyle 1944) could power a
significant portion of the
observed X-rays ($L_{38}\approx0.3$).  However, we can not rule out that
the low velocity and high wind density we have measured are attributable
not to an isotropic wind, but to other gas in the system, such as the gas
stream or a wind-compressed disk about the primary. 

\acknowledgements

We would like to thank M. Preciado for his assistance.  This work was
based on observations with the NASA/ESA {\it Hubble Space Telescope},
obtained at the Space Telescope Science Institute, which is operated by
the Association of Universities for Research in Astronomy, Inc., under
NASA contract GO-05874.01-94A.  BB and SDV supported in part by NASA
(NAG5-2532, NAGW-2685), and NSF (DGE-9350074).  BB acknowledges an NRC
postdoctoral associateship.

\clearpage

\figcaption{A comparison of the LMC X-4 \ion{N}{5} line
profiles with
those of Vela X-1 and 4U1700-37.}

\figcaption{Comparison of the \ion{N}{5} lines in LMC~X-4 observed at
$\phi=0.42$ with (a) the same region in the spectrum of $\mu$Col
(dot-dashed; the flux and systemic velocity have been normalized to that
of LMC~X-4), and (b) Gaussian lines at the systemic velocity of the LMC
and with the same width as optical lines attributed to the photosphere of
the primary. The vertical lines show the rest wavelenths of the \ion{N}{5}
lines.}

\figcaption{The geometry of the Hatchett-McCray $q$ parameter.  Surfaces
with constant values of $q$ bound regions of the wind at different
ionization levels.  The primary star is represented as a circle centered
at the origin, and the X-ray source is marked with an `X'.}

\figcaption{Variations in the equivalent width of a) \ion{N}{5}, b)
\ion{C}{4}, and
c) \ion{Si}{4}. The ``*'' signs mark the equivalent widths from IUE
observations reported by van der Klis et al. (1982), while the ``+'' signs
are based on the data reported in Paper~I.  The error bars show the
$1\sigma$ error range.  Also shown are theoretical
models in which the ion is removed from regions of the wind with various
values of the Hatchett-McCray parameter $q$.}

\figcaption{The geometry used by a simple analytic method to infer the 
radial wind velocity from P~Cygni line variations.  The viewer is towards 
the bottom of the page.  The circle represents the surface of the primary star
and the point marked ``NS'' represents the neutron star.  $V(\phi)$ marks
the highest observed velocity of absorption.  In (a) we show the
situation at $\phi=0.21$ and in (b) we show $\phi=0.32$.}

\figcaption{An example of the measurement of $V(\phi)$, the maximum velocity
of absorption at a given orbital phase.  We fit the edge of the observed line
profile (the dotted line shows the smoothed spectrum) to a horizontal line
segment for blue-shifts $V>V(\phi)$ and to a line segment with slope as a
free parameter for $V<V(\phi)$.  The vertical line marks the measured value
of $V(\phi)$ for the 1242.8\AA\ doublet component, determined by where the
continuum reaches the same level as for the horizontal line segment.}

\figcaption{The wind expansion velocity versus radial distance, as
inferred from the P~Cygni line profiles.  The ``*'' signs mark points
determined from the absorption profile.  For points marked with ``+''
signs, a model of the blue-shifted emission as the red-shifted emission 
reflected about $v=150$~km~s$^{-1}$ has been subtracted from the
P~Cygni profiles.  For points marked with diamonds, the red-shifted
emission was reflected about $v=0$~km~s$^{-1}$.
Points marked with ``x'' were determined from differences of spectra
corrected for blue-shifted emission. The solid line shows the best fit
for the diamonds for $\phi<0.25$ ($R>2$).}

\figcaption{An illustration of a model that fits the absorption line
profiles to determine the wind acceleration parameter $\beta$.  The
conical X-ray shadow of the primary intersects isovelocity surfaces at
$v/v_{\infty}=0.2,0.4$.  These intersections determine images of the
primary's surface at Doppler-shifted wavelengths corresponding to
$v/v_{\infty}=0.2,0.4$.  The contribution to the absorption line profile
at $v/v_{\infty}=0.2,0.4$ is determined by the fraction of the primary's
surface that is not in shadow.}

\figcaption{The best-fit model (dashed) to the \ion{N}{5} absorption line
profiles, allowing the wind acceleration parameter $\beta$ and the wind
optical depth $\tau$ as free parameters, and assuming the X-rays
photoionize the entire wind outside the shadow of the primary.  We have
subtracted a model of the blue-shifted P~Cygni emission in order to fit
only the blue-shifted P~Cygni absorption.}

\figcaption{The results of the best-fit model of the P~Cygni line variations.
The right panel shows the differences between successive orbital phases,
for both the data and model.}

\figcaption{The results of the best-fit model of the P~Cygni line 
variations, for the inhomogeneous wind model.  The right-hand panels show
the differences between successive spectra.}

\figcaption{The results of the best-fit model of the \ion{C}{4} P~Cygni
line
variations.}



\clearpage

\begin{deluxetable}{ccrrrrrr}
\tablewidth{33pc}
\tablehead{
\colhead{$\phi$} & \colhead{$\phi^\prime$} & \colhead{$V_{\rm
obs}$\tablenotemark{a}} & \colhead{$V_{\rm
sub,0}$\tablenotemark{b}} 
& \colhead{$V_{\rm max,0}$\tablenotemark{c}} & \colhead{$D_2$} & 
\colhead{$D_3$} & \colhead{$D_4$}\nl
       &               & (km/s)   & (km/s) &   (km/s) & (R$_\star$)
& (R$_\star$) & (R$_\star$)}
\startdata
0.095 & 0.114 & 1270 & 1260 & 1260 & 8.82 & 17.57 & 
17.60\\ 
0.099 & 0.117 & 1450 & 1390 & 1390 & 7.59   & 15.11 &  15.15\\
0.102 & 0.120 & 1220 & 1210 & 1210 & 6.66   & 13.24  &  13.27\\
0.106 & 0.123 & 1250 & 1200 & 1210 & 5.92   & 11.76  &   11.80\\
0.110 & 0.126 & 1170 & 1100 & 1100 & 5.33   & 10.56  &  10.61\\
0.114 & 0.129 & 1210 & 1190 & 1190 & 4.84   & 9.58  &  9.63\\
0.117 & 0.132 & 1270 & 1100 & 1110 & 4.43   & 8.76 &  8.82\\
0.121 & 0.135 & 1210 & 1130 & 1130 & 4.09  & 8.06 &  8.12\\
0.125 & 0.138 & 1140 & 1110 & 1120 &  3.79  & 7.45 &  7.52\\
0.128 & 0.141 & 1250 & 1110 & 1120 & 3.54  & 6.93 &  7.00\\
\tableline
0.195 & 0.200 & 490,410 & 820,720   & 810 & 1.64 & 2.94 & 
3.11\\
0.199 & 0.203 & 470,350 & 750,700       & 770 & 1.60 & 2.84 & 3.01\\
0.202 & 0.207 & 430,460 & 810,740       & 820 & 1.56 & 2.75 & 2.93\\
0.206 & 0.210 & 450,350 & 660,720       & 740 & 1.52 & 2.66 & 2.85\\
0.210 & 0.213 & 380,340 & 750,690       & 770 & 1.48 & 2.58 & 2.77\\
0.213 & 0.217 & 410,480 & 640,700       & 720 & 1.45 & 2.50 & 2.69\\
0.217 & 0.220 & 360,390 & 630,700       & 720 & 1.42 & 2.43 & 2.63\\
0.221 & 0.223 & 450,340 & 620,720       & 730 & 1.39 & 2.36 & 2.56\\
0.224 & 0.227 & 360,370 & 580,700       & 700 & 1.36 & 2.29 & 2.50\\
0.228 & 0.230 & 390,320 & 590,650       & 680 & 1.34 & 2.22 & 2.44\\
\tableline
\tablebreak
0.293 & 0.289 & 350,220 & 430,280 & 430 & 1.07 & 1.43 & 1.75\\
0.297 & 0.293 & 370,190 & 520,380 & 560 & 1.06 & 1.40 & 1.72\\
0.300 & 0.296 & 300,260 & 490,470       & 590 & 1.05 & 1.37 & 1.70\\
0.304 & 0.299 & 280,260 & 280,350       & 390 & 1.04 & 1.34 & 1.68\\
0.308 & 0.302 & 300,260 & 460,220       & 430 & 1.04 & 1.32 & 1.65\\
0.311 & 0.306 & 390,210 & 520,190       & 450 & 1.03 & 1.29 & 1.63\\
0.315 & 0.309 & 320,240 & 410,220       & 400 & 1.03 & 1.26 & 1.61\\
0.319 & 0.312 & 340,210 & 460,170       & 400 & 1.02 & 1.23 & 1.59\\
0.322 & 0.316 & 340,260 & 470,310       & 510 & 1.02 & 1.21 & 1.57\\
0.326 & 0.319 & 350,220 & 560,400       & 630 & 1.01 & 1.18 & 1.55\\
\enddata
\tablenotetext{a}{The maximum observed absorption velocity in the
uncorrected \ion{N}{5} line profile; where two numbers are given these are
the blue and red doublet components, respectively.  All velocities
are relative to 
the LMC (at $+280$~km~s$^{-1}$)}
\tablenotetext{b}{The maximum observed absorption velocity in the line
profile corrected for blue-shifted emission}
\tablenotetext{c}{The maximum radial wind velocity based on the observed
line-of-sight velocity}
\tablecaption{Measurements of the Wind Radial Velocity in LMC~X-4}
\end{deluxetable}

\clearpage

\begin{table*}
\begin{center}
\begin{tabular}{ccccc}
$\chi^2$ & $v_\infty$ & $\beta$ & $v_0$ & v$_{\rm subtract}$\\
 & (km s$^{-1}$) & & (km s$^{-1}$) &\\
\hline
2.79 & $1800\pm200$ & $2.3\pm0.6$ & $200\pm200$ & NA \\
0.89 & $1320\pm40$ & $1.39\pm0.14$ & $-40\pm50$ & 0 \\
1.94 & $1400\pm100$ & $2.4\pm0.8$ & $-140\pm130$ & 150 \
\end{tabular}
\end{center}
\caption{Analytic fits to the wind radial-velocity law}
\end{table*}

\clearpage

\begin{table*}
\begin{center}
\begin{tabular}{lccc}
Parameter & Meaning & Value &  Value\\
          &         & (basic model)& (inhomogeneous wind)\\
\tableline
T & Background wind optical depth maximum & 1.13$\pm0.01$ &
1.00$\pm0.03,700\pm200$\\
$\gamma$ & Background wind optical depth exponent & 0.77$\pm0.04$ & 
2.1$\pm0.1$\\
$v_{\infty}$ & Wind terminal velocity & 1350$\pm35$ & 1280$\pm35$ \\ 
$v_0$ & Velocity relative LMC & 80$\pm35$ & 15$\pm35$ \\
$L_{38}/\bar{dot{M_{-6}}}$ & X-ray luminosity/Mass-loss rate & 
$0.26\pm0.01$ & 0.7$\pm0.4$ \\
$\tau_{\rm b}$ & Optical depth of photospheric line (blue) & 
$0.36\pm0.01$ & $0.62\pm0.02$\\
$\chi^2_\nu$, dof & Goodness of fit, degrees of freedom & 5.7, 690 & 4.0,
688\\
\end{tabular}
\end{center}
\caption{Best-fit parameters for the numeric model of the \ion{N}{5} 
P~Cygni 
profile variations} \vspace{0.5in}
\end{table*}

\clearpage

\setcounter{figure}{0}
\begin{figure}
\caption{}
\plotone{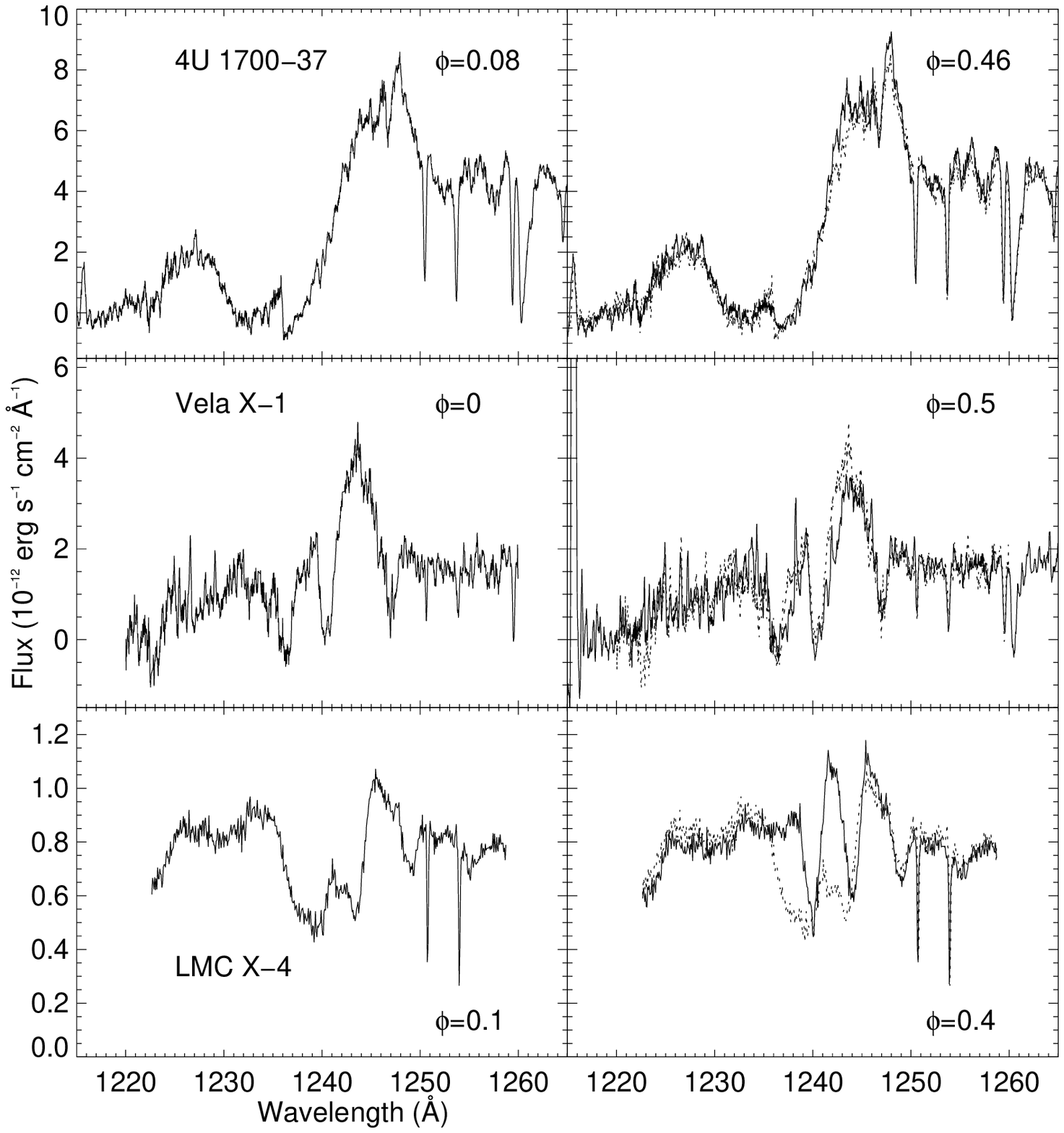}
\end{figure}

\begin{figure}[h]
\caption{}
\plotone{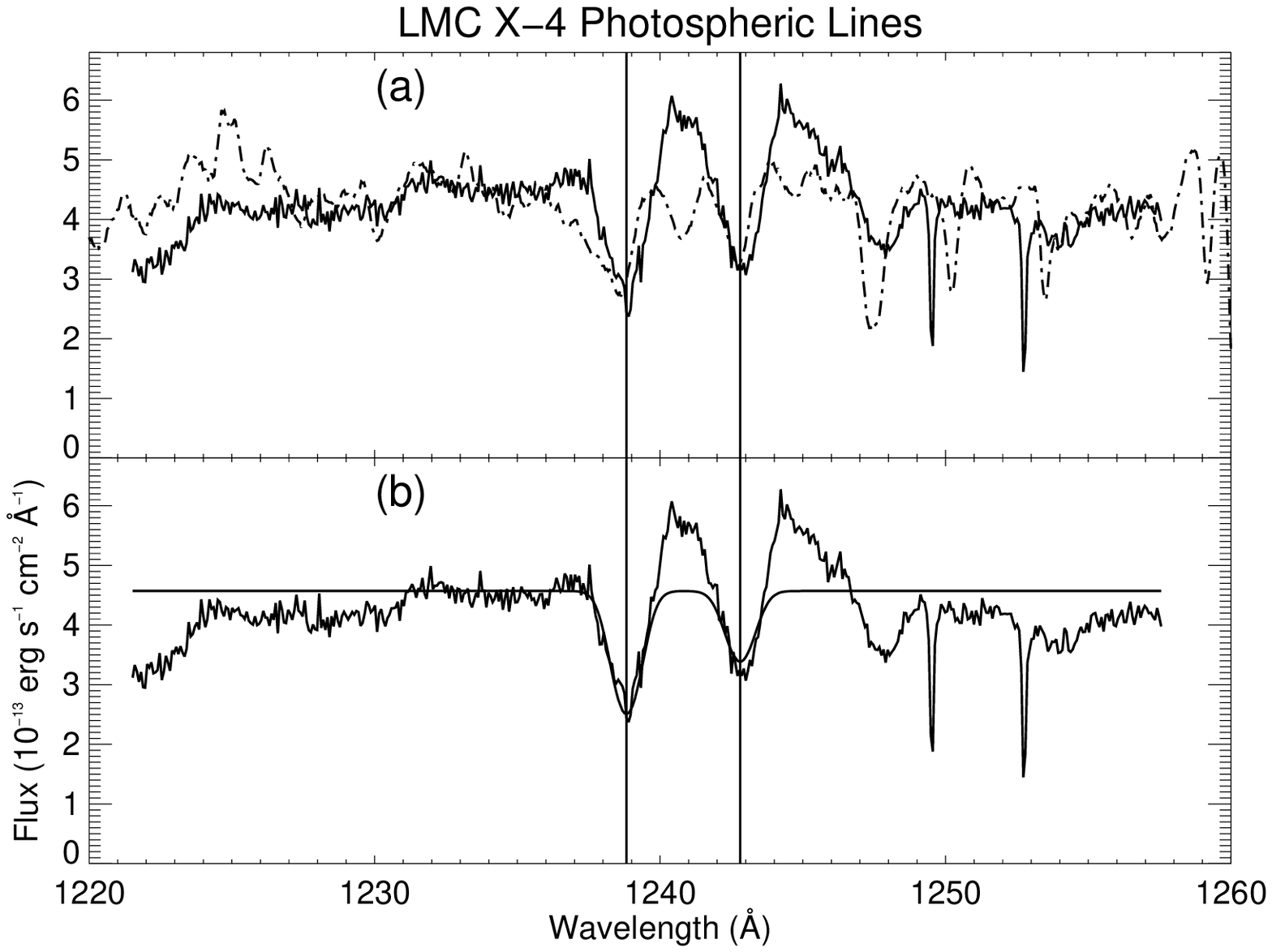}
\end{figure}

\begin{figure}
\caption{}
\plotone{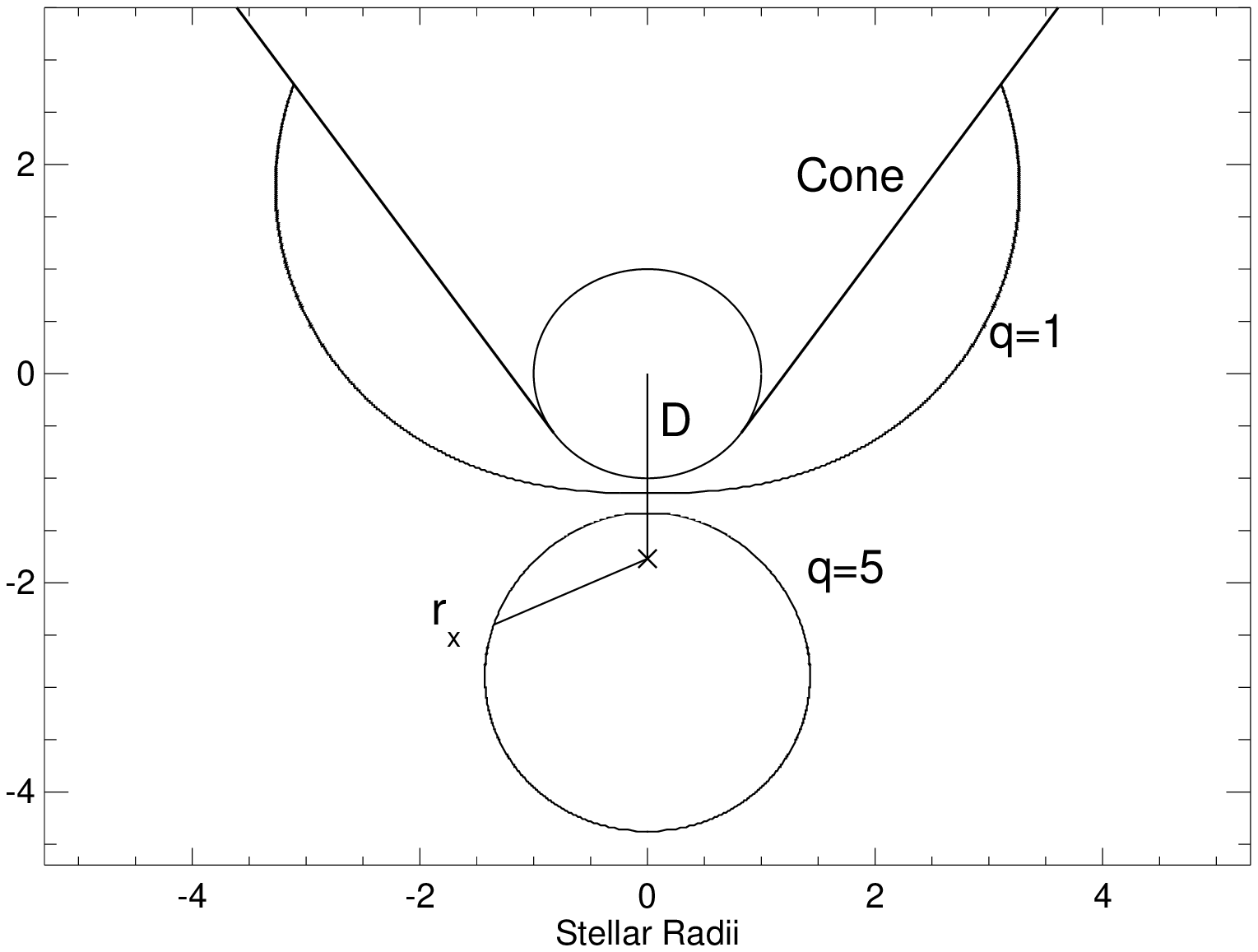}
\end{figure}

\begin{figure}
\caption{}
\plotone{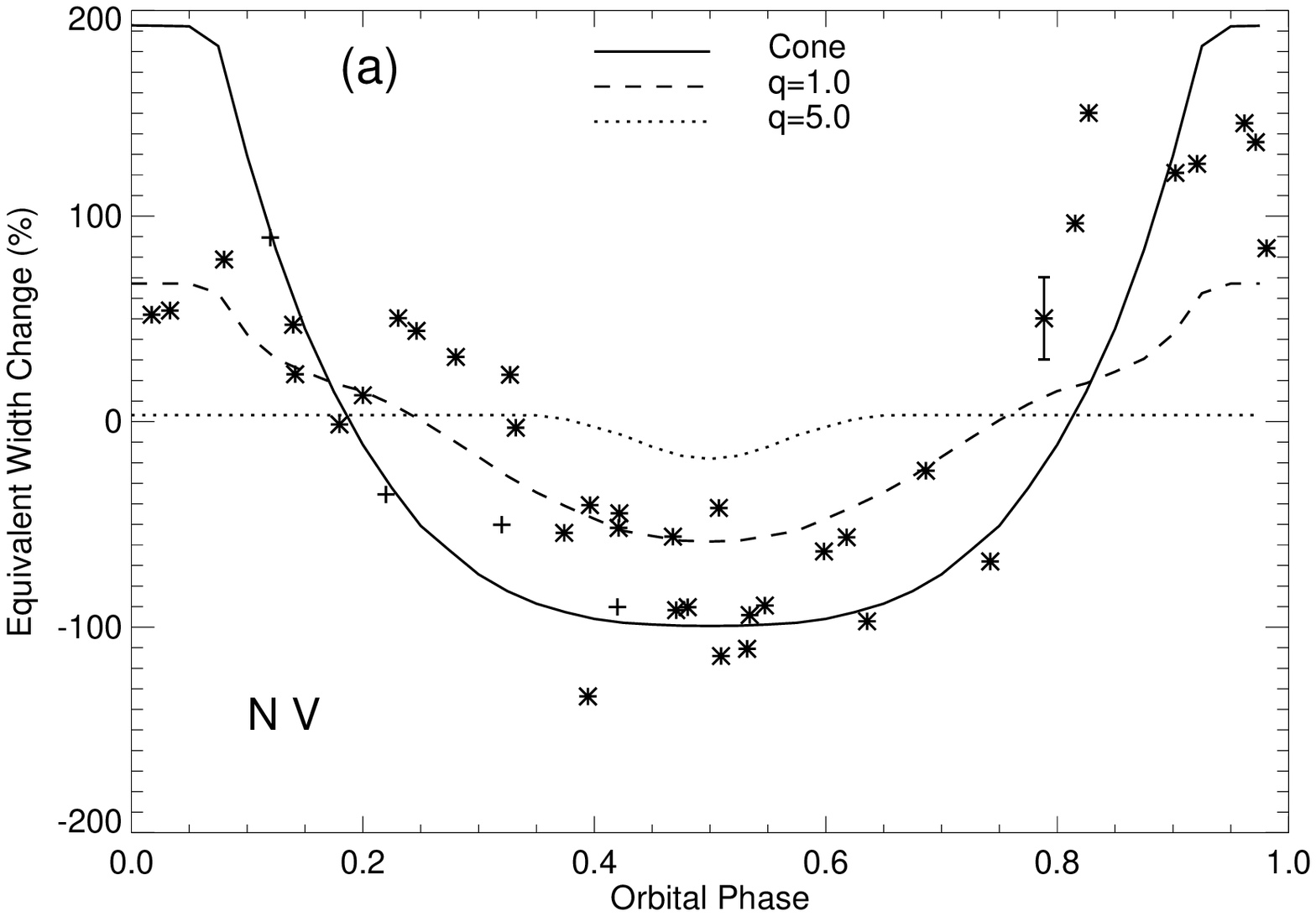}
\end{figure}

\setcounter{figure}{3}
\begin{figure}
\plotone{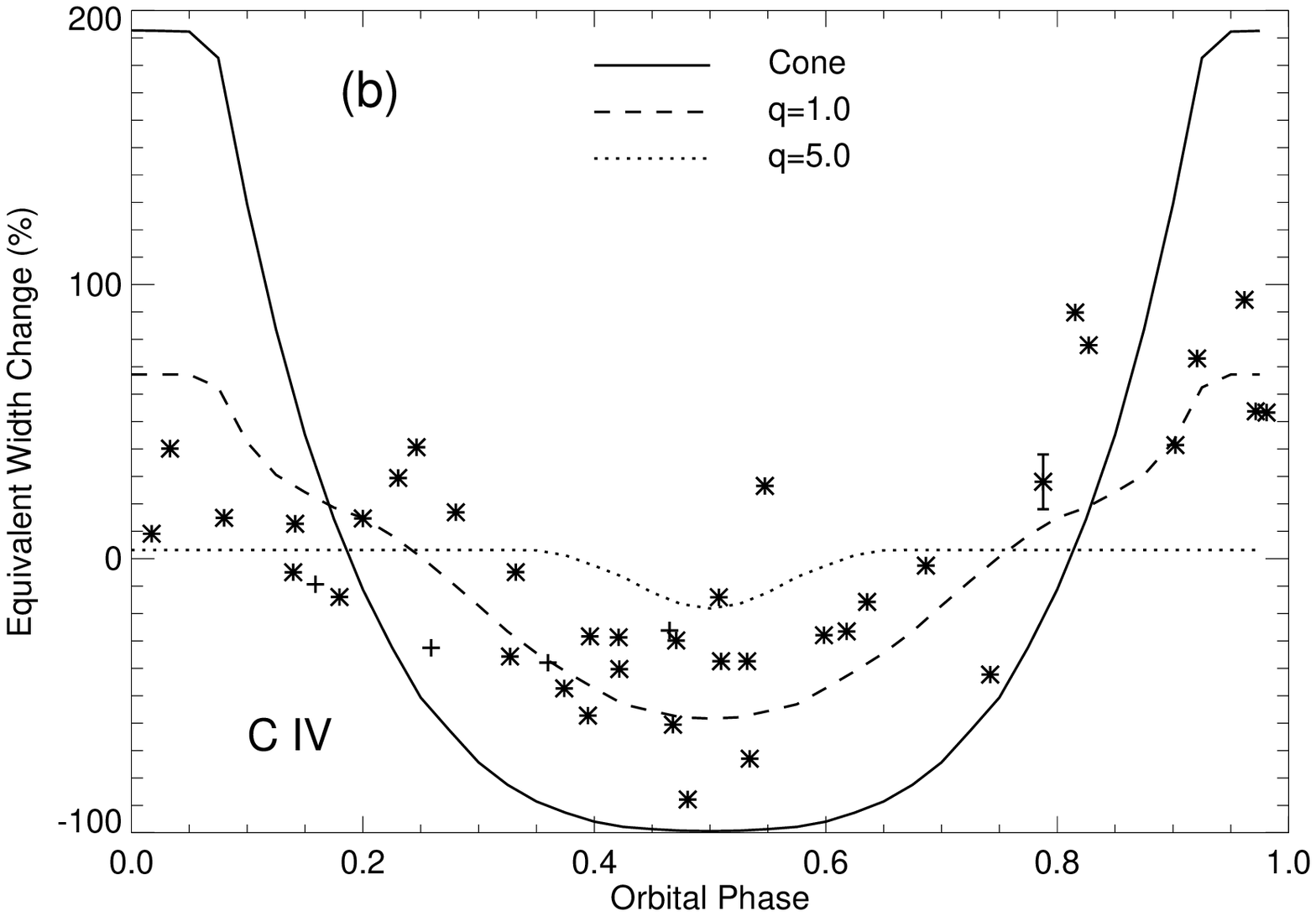}
\end{figure}

\setcounter{figure}{3}
\begin{figure}
\caption{}
\plotone{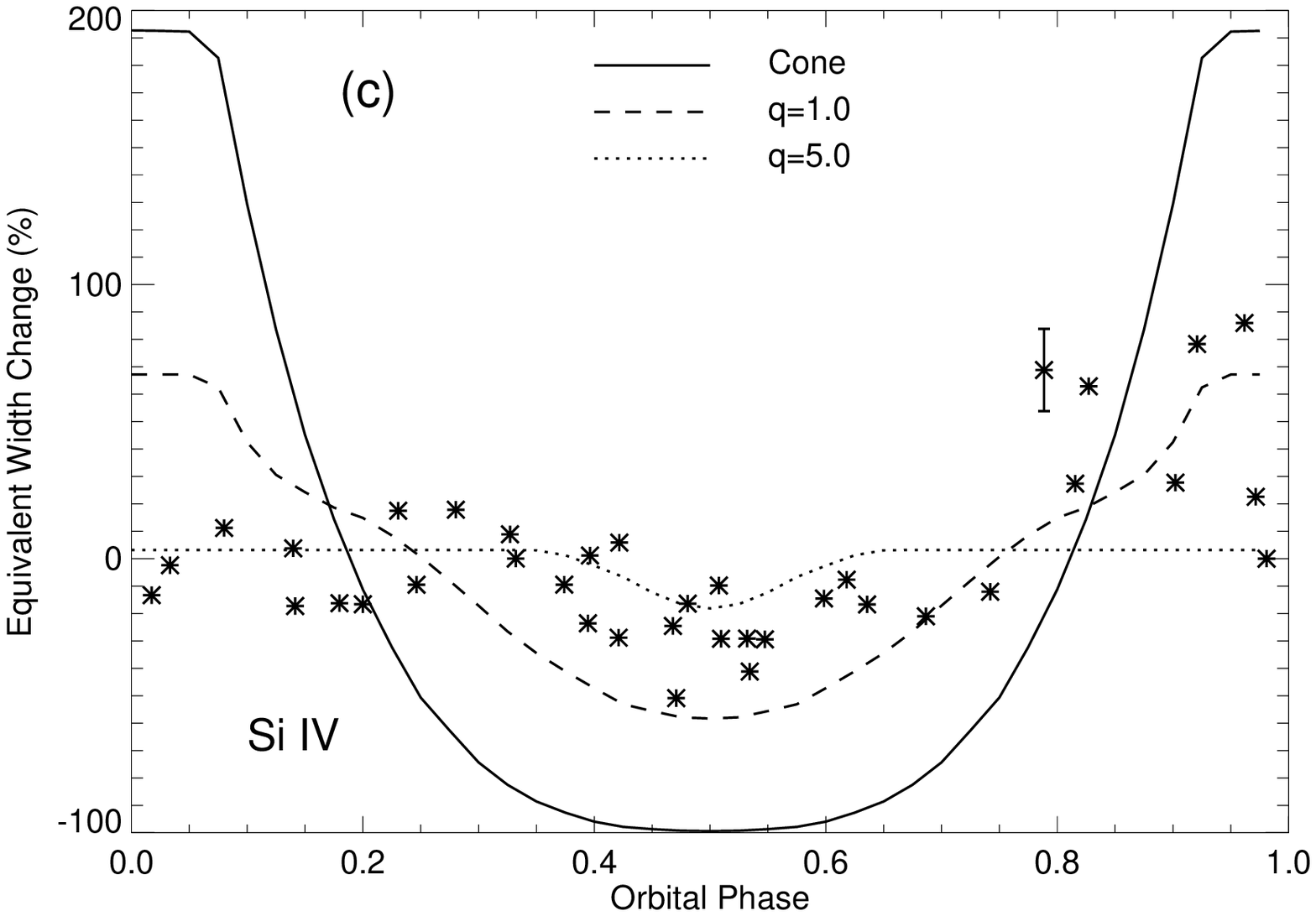}
\end{figure}

\begin{figure}
\caption{}
\plotone{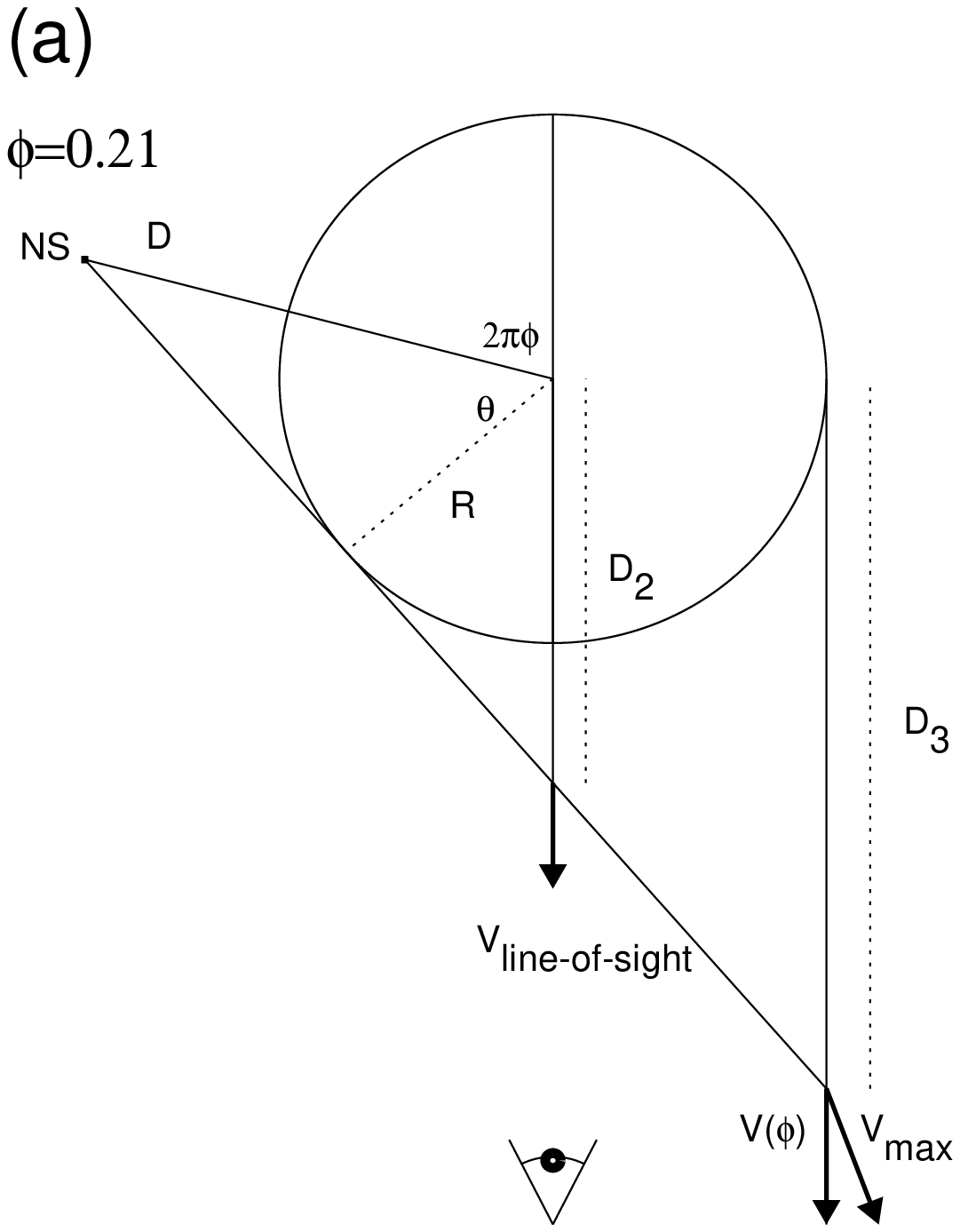}
\end{figure}

\setcounter{figure}{4}
\begin{figure}
\caption{}
\plotone{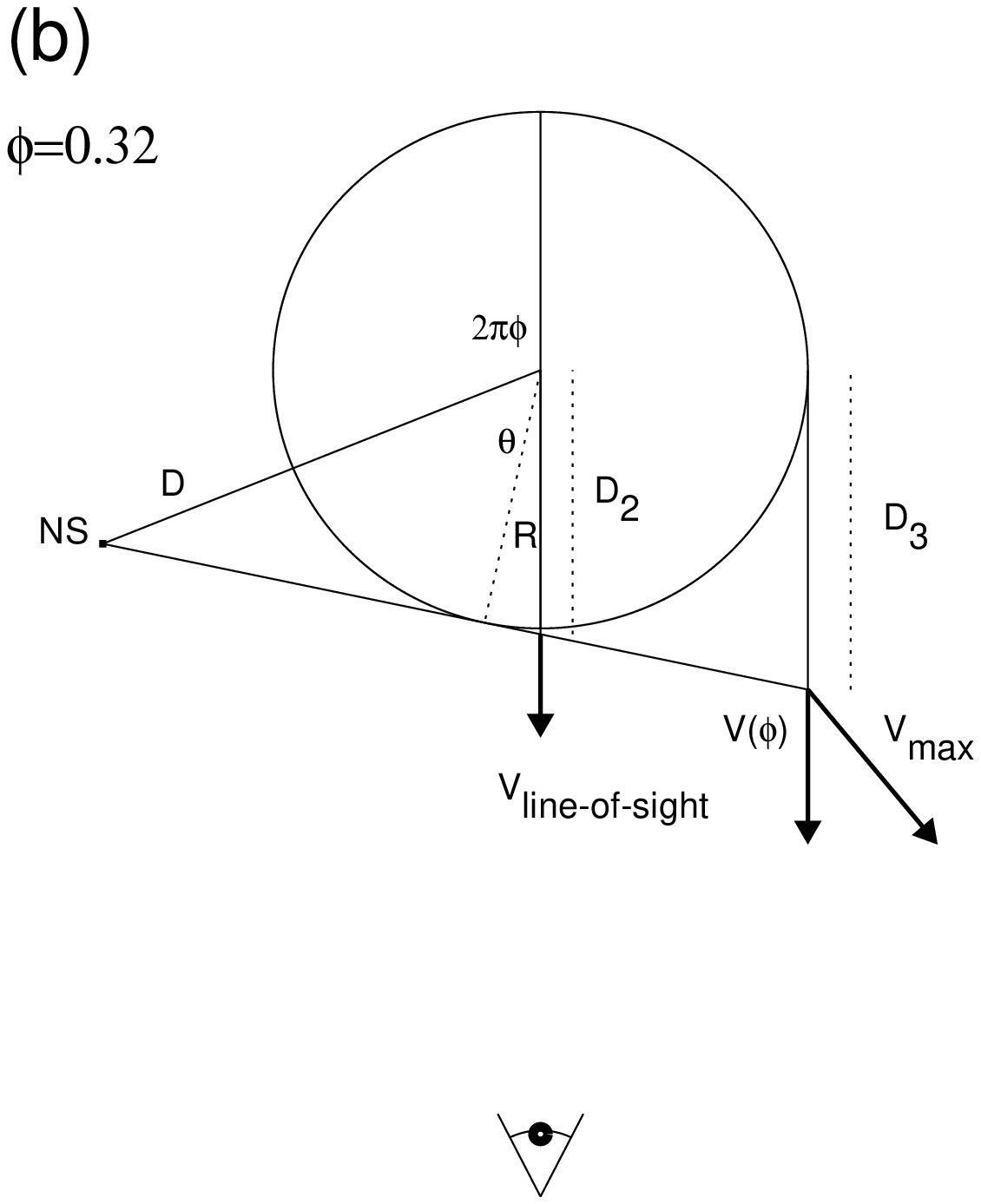}
\end{figure}

\begin{figure}
\caption{}
\plotone{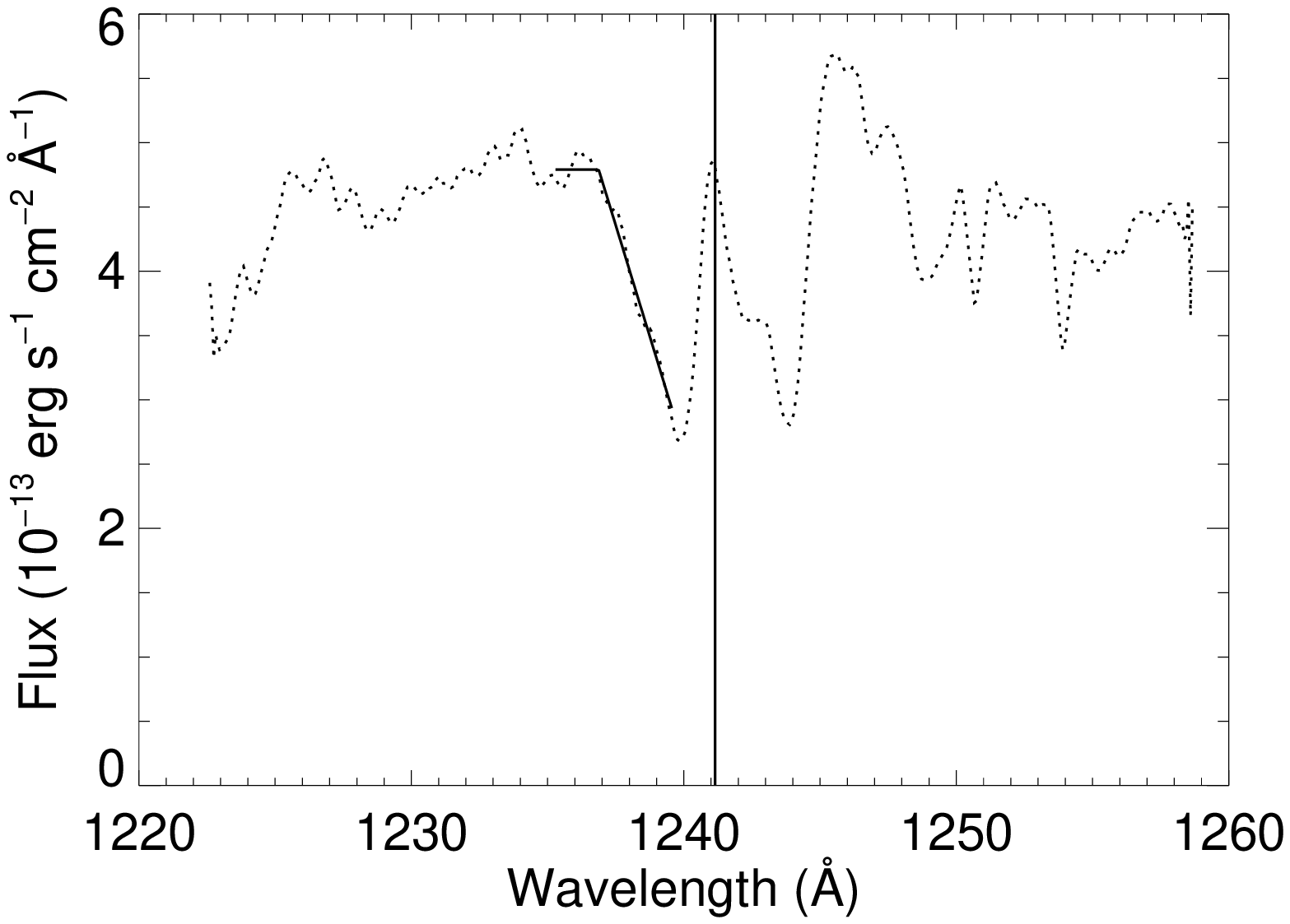}
\end{figure}
\clearpage

\begin{figure}
\caption{}
\plotone{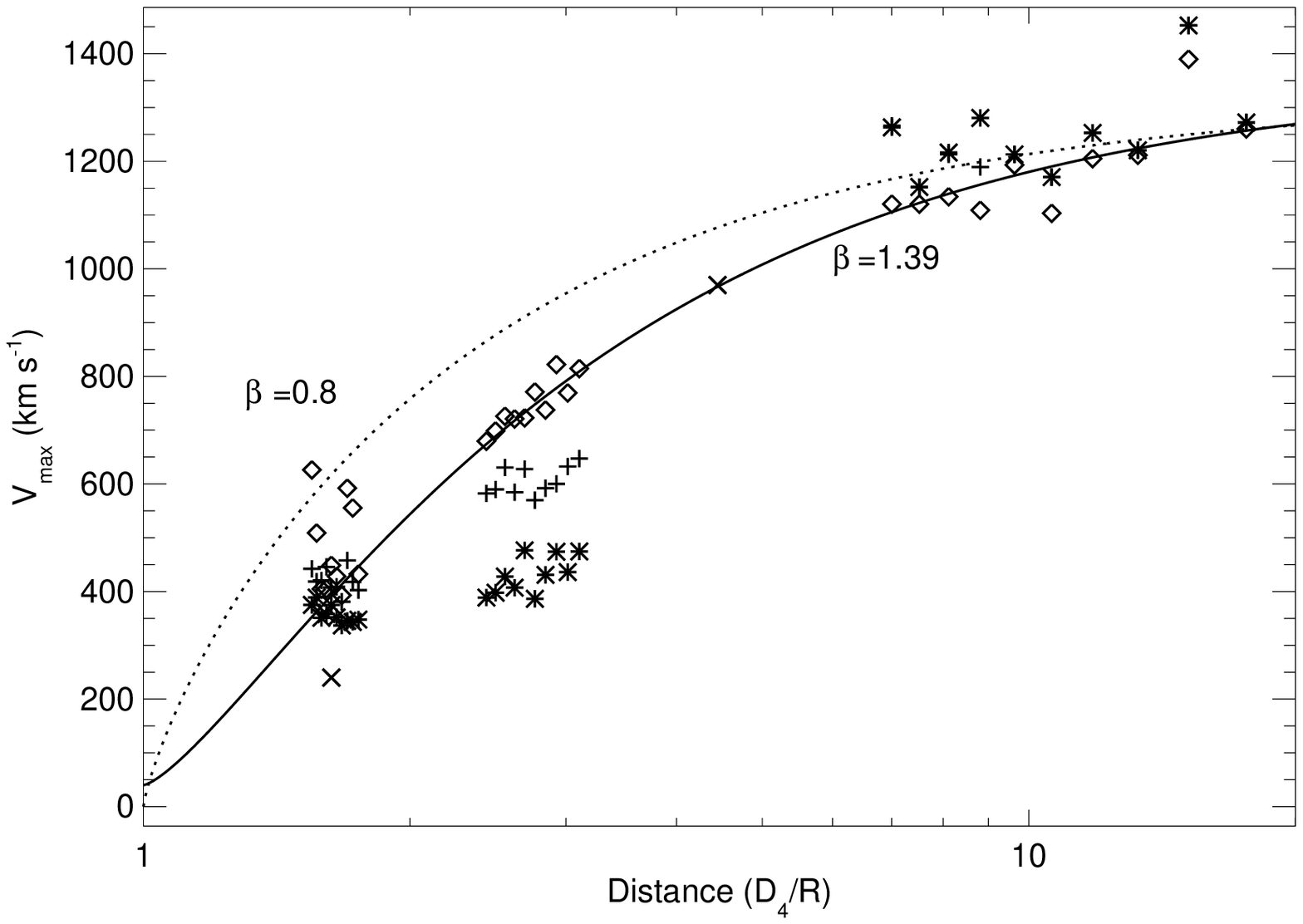}
\end{figure}
\clearpage

\begin{figure}
\caption{}
\plotone{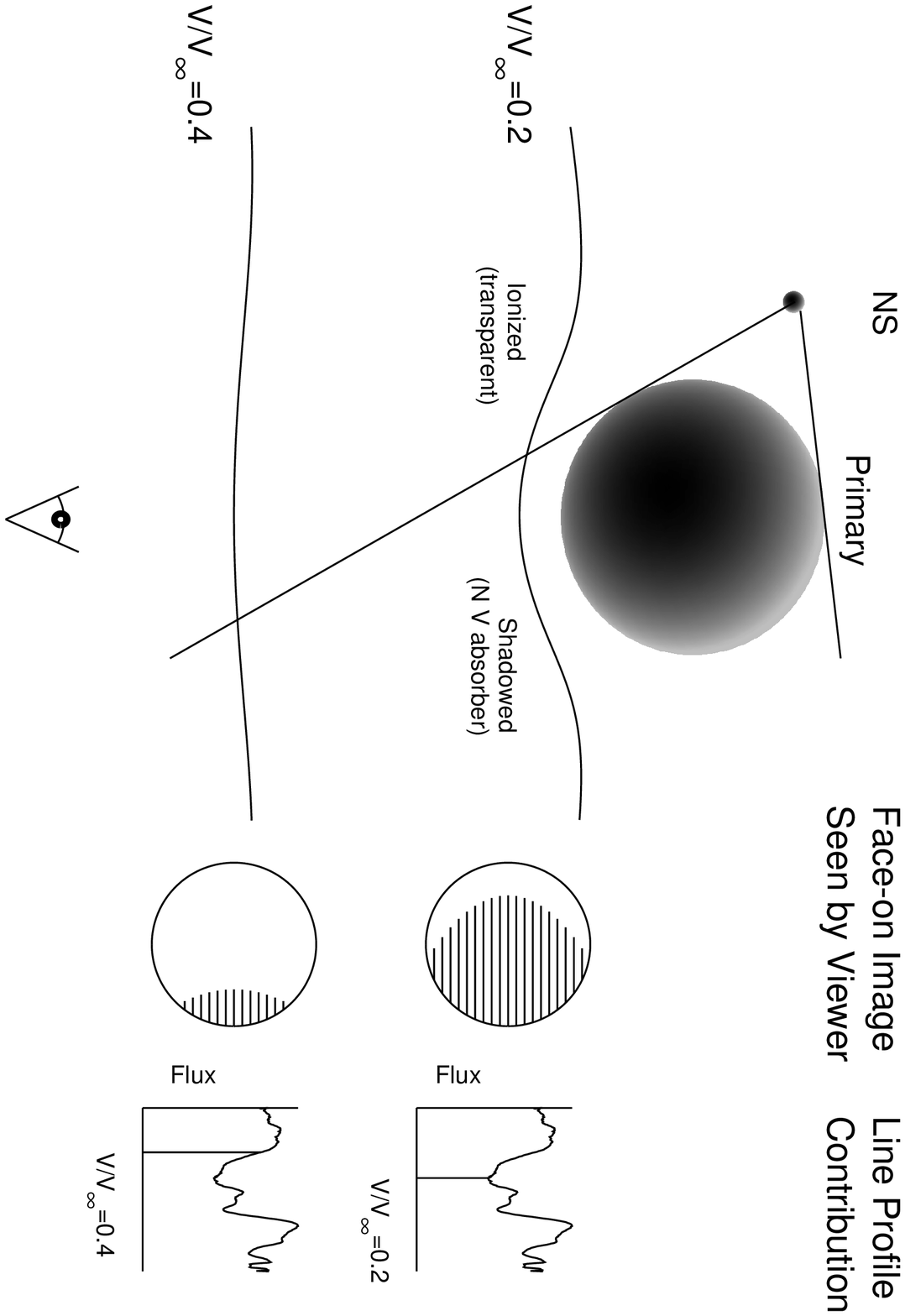}
\end{figure}
\clearpage

\begin{figure}
\caption{}
\plotone{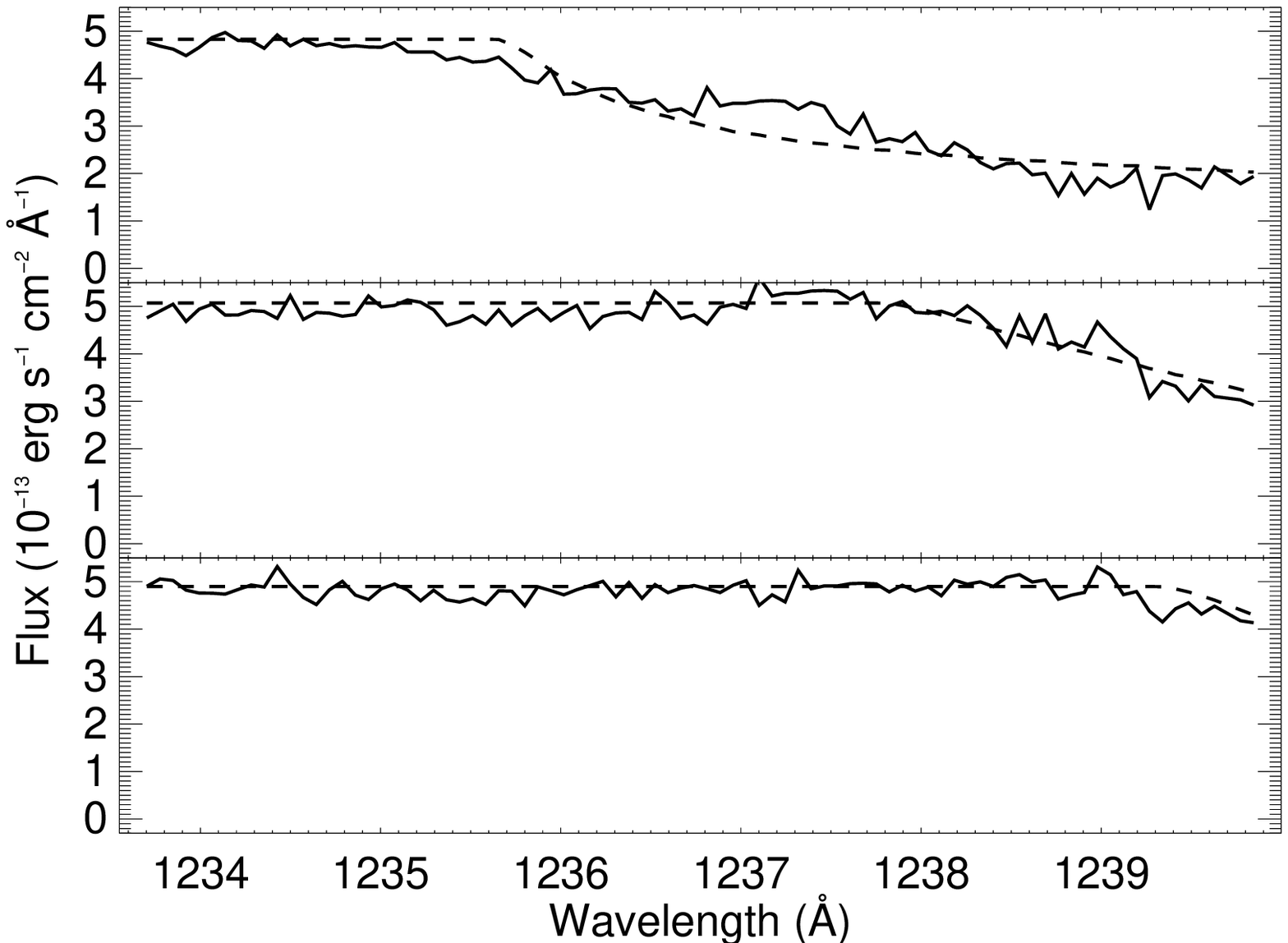}
\end{figure}

\begin{figure}
\caption{}
\plotone{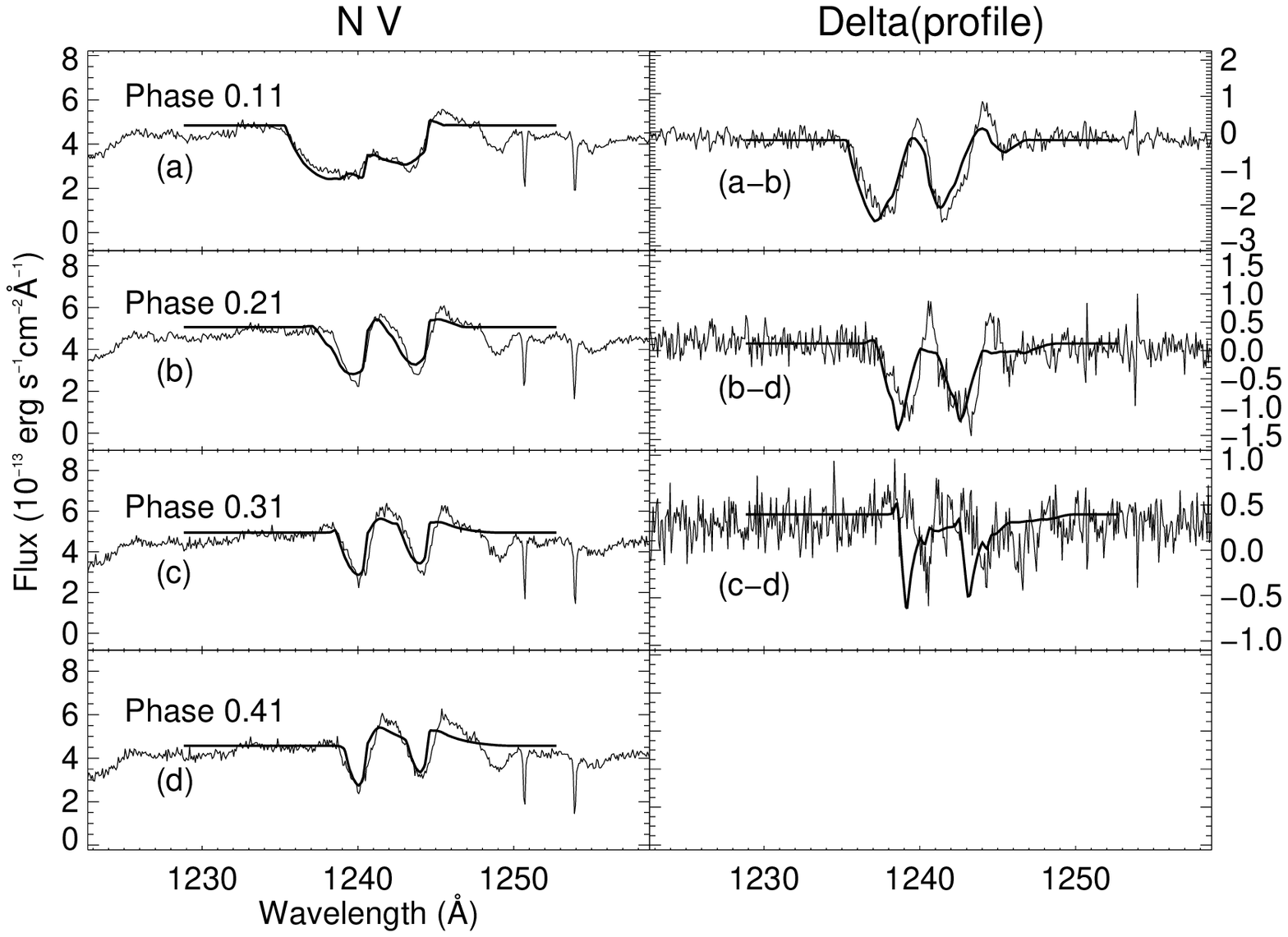}
\end{figure}

\begin{figure}
\caption{}
\plotone{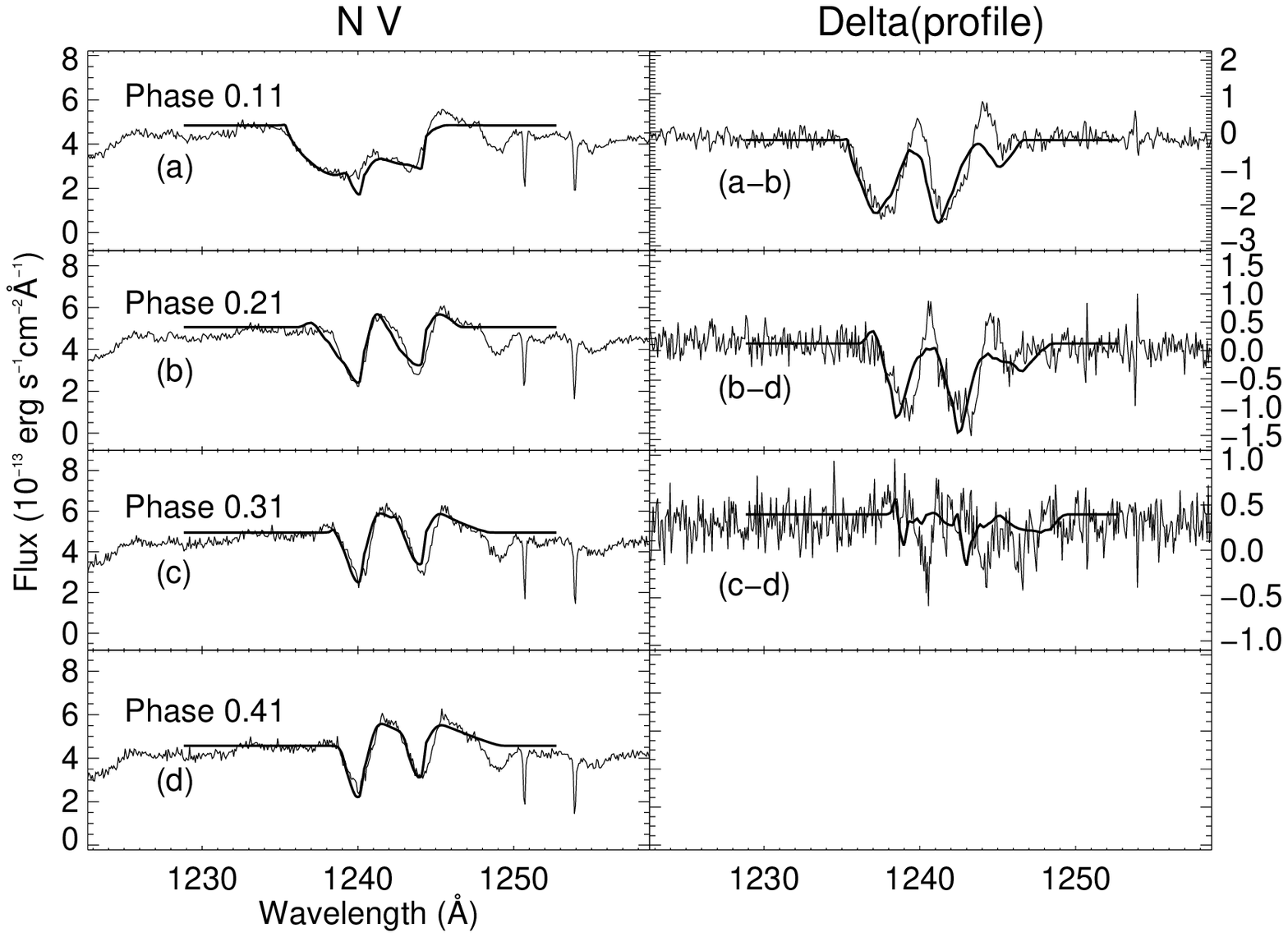}
\end{figure}

\begin{figure}
\caption{}
\plotone{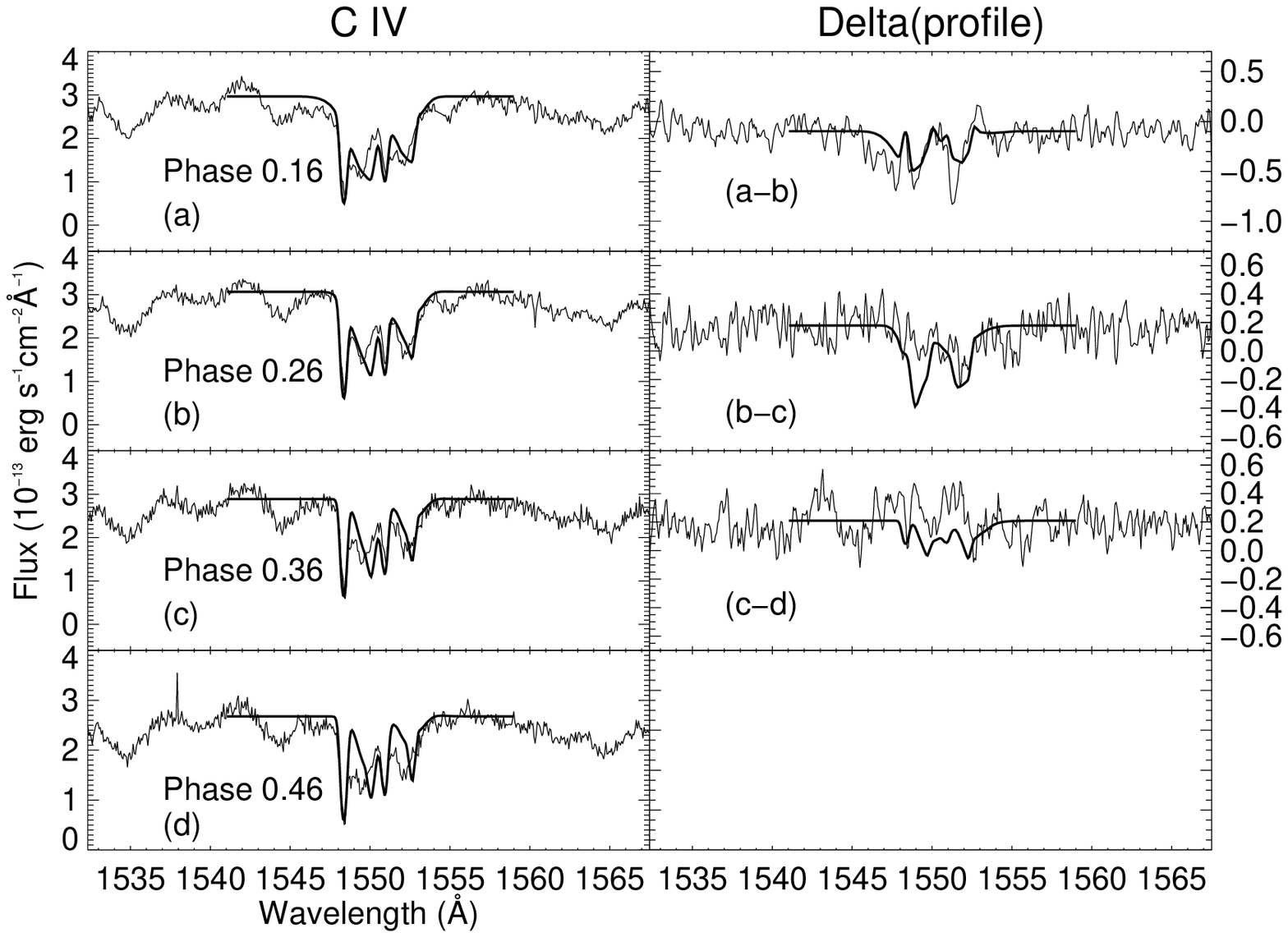}
\end{figure}



\end{document}